\documentclass[lettersize,journal]{IEEEtran}

\usepackage{wrapfig}
\usepackage[utf8]{inputenc}
\usepackage{algpseudocode}
\usepackage{amsmath,amsfonts}
\usepackage{algorithm}
\usepackage{array}
\usepackage[caption=false,font=normalsize,labelfont=sf,textfont=sf]{subfig}
\usepackage{textcomp}
\usepackage{stfloats}
\usepackage{url}
\usepackage{verbatim}
\usepackage{tabularx}
\usepackage{booktabs}
\usepackage{pifont}
\usepackage{graphicx}
\usepackage{cite}
\usepackage{tikz,xcolor}
\usepackage{amssymb}
\usepackage{makecell}
\DeclareRobustCommand\encircle[1]{%
  \tikz[baseline=(c.base)]{
    \node[
      shape=circle,
      draw,
      fill=white,
      inner sep=0.16ex,    
      minimum size=1.9ex,  
      line width=0.35pt
    ] (c) {\fontsize{7.5}{7.5}\selectfont #1};
  }%
}

\begin{document}

\title{GEN-Graph: Heterogeneous PIM Accelerator for General Computational Patterns in Graph-based Dynamic Programming}

\author{
    \IEEEauthorblockN{
        Yanru Chen\textsuperscript{1}, 
        Runyang Tian\textsuperscript{1}, Zheyu Li\textsuperscript{1}, Mahbod Afarin\textsuperscript{1}, 
        Weihong Xu\textsuperscript{2}, 
        Tajana Šimunić Rosing\textsuperscript{1}~\IEEEmembership{Fellow,~IEEE}
    }

    \IEEEauthorblockA{
        \textsuperscript{1}University of California San Diego, La Jolla, CA, USA\\
    }

    \IEEEauthorblockA{
        \textsuperscript{2}Ecole Polytechnique Fédérale de Lausanne, Lausanne, Switzerland\\
    }

    \IEEEauthorblockA{
        \{yac054, r3tian, zhl178, mafarin, tajana\}@ucsd.edu; weihong.xu@epfl.ch
    }

}

\markboth{Journal of \LaTeX\ Class Files,~Vol.~14, No.~8, August~2021}%
{Shell \MakeLowercase{\textit{et al.}}: A Sample Article Using IEEEtran.cls for IEEE Journals}

\IEEEaftertitletext{\vspace{-0.8cm}}

\maketitle

\begin{abstract}
While graph-based dynamic programming (DP) is a cornerstone of genomics and network analytics, its efficiency is hampered by fundamentally conflicting computational patterns. Matrix-centric DP drives regular, compute-bound network analytics, while topology-centric DP handles irregular, memory-bound genomic traversals. These two categories of DP have substantially different computation patterns and dataflows, which makes it difficult for a single homogeneous processing‑in‑memory (PIM) architecture to efficiently support both.
This work presents GEN‑Graph, a novel heterogeneous PIM chiplet that integrates two types of specialized compute tiles within a 2.5D package: Matrix-tile, a processing‑using‑memory (PUM) tile optimized for matrix-centric workloads, such as all-pairs shortest path (APSP); and traversal-tile, a processing‑near‑memory (PNM) tile optimized for traversal-centric DP workloads, such as DNA sequence alignment. Our hardware-software co-design employs recursive partitioning and reconfigurable windowed bit-parallel logic to ensure exact computation. Results show the matrix tile achieves $42.8\times$ speedup and $392\times$ energy efficiency over the NVIDIA H100 GPU for APSP. For sequence-to-graph alignment, the traversal tile sustains $2.56$ million reads/s (short-reads) and $39.3$ thousand reads/s (long-reads), outperforming state-of-the-art accelerators by up to $2.56\times$ in throughput. GEN-Graph provides the first scalable, exact solution for general DP dataflows by matching hardware specialization to algorithmic structure.

\end{abstract}


\section{Introduction}

Graph analytics provides the computational foundation for a vast range of applications spanning social network analysis~\cite{rajaraman2011mining,ma2024using,simas2021distance}, bioinformatics~\cite{paten2017genome,ameur2019goodbye,rakocevic2019fast,eizenga2020pangenome}, and logistics~\cite{cattaruzza2017vehicle,wan2021survey}. As social networks~\cite{snap_large_networks} and genomics graphs~\cite{jain2018nanopore,zook2016extensive} scale beyond billions of nodes, they demand different algorithmic strategies. These range from fundamental traversals like Breadth-First Search (BFS)~\cite{murphy2010introducing} for connectivity, to iterative methods like PageRank~\cite{page1999pagerank} focused on node importance. Dynamic programming (DP)~\cite{bellman1966dynamic,cormen2022introduction} further optimizes complex path-finding by memoizing overlapping subproblems. By avoiding full-graph recomputation, DP enables efficient, real-time updates in large-scale evolving networks—a critical capability across numerous domains. In bioinformatics, key applications include sequence-to-graph (S2G) alignment~\cite{garrison2018variation}, which is crucial for modern pangenome analysis~\cite{liao2023draft}. In broader domains such as network science and transportation, the all-pairs shortest path (APSP)~\cite{floyd1962algorithm} problem remains a classic and vital computation.

Although these algorithms share mathematical foundations, they employ diverse computational structures with unique recurrence relations, data dependencies, and update mechanisms~\cite{mcsherry2015cost,kepner2011graph,shun2013ligra}. We classify graph DP into two distinct computational archetypes: matrix-centric DP and topology-centric DP. Matrix-centric DP problems operate on static graph structures where connectivity is dense or can be regularized. The dataflow resembles blocked matrix multiplication, characterized by high spatial locality and predictable reuse~\cite{goto2008anatomy,williams2009roofline}. Here, the bottleneck is peak arithmetic throughput. Conversely, topology-centric DP kernels exhibit large-scale graphs where edges represent dynamic transitions~\cite{shun2013ligra}. The dataflow is dominated by pointer chasing and indirect memory addressing. This poor data locality leads to frequent memory stalls and low effective bandwidth from random memory access. These distinct dataflows create a significant gap in general graph-based DP hardware acceleration. To understand how to bridge this gap, we analyze APSP and S2G alignment as representative workloads for these two archetypes in Section~\ref{sec:roofline}.

Recent GPU-based accelerators launch massive parallelism, but graph DP still faces scaling and efficiency limits~\cite{shi2018graph,xie2025gpu}. For instance, the performance of dense APSP is dictated by the efficiency of executing $O(n^3)$ min-plus (MP) matrix multiplication operations~\cite{djidjev2015all}. State-of-the-art (SOTA) designs focus on offloading this compute-intensive part to massively parallel GPU architectures to approach theoretical peak performance: Partitioned-APSP computes APSP for a 2M-vertex graph in approximately 30 minutes but requires 128 GPUs with extensive DRAM reliance~\cite{djidjev2015all}; Co-ParallelFW achieves 8.1\,PFLOP/s but requires complex coordination among 4\,608 GPUs~\cite{sao2021scalable}, showing the need for large hardware footprints, heavy interconnect communication, and high energy~\cite{shi2018graph,xie2025gpu}. In SOTA traversal sequence alignment accelerator designs, heterogeneous graph aligner (HGA)~\cite{feng2021accelerating} achieves a $15.8\times$ speedup via CPU-GPU co-processing, yet it remains constrained by long reads where larger alignment scores necessitate more bits, thereby reducing SIMD lanes and overall parallel efficiency. Overall, data movement remains a dominant bottleneck for graph DP at scale~\cite{wulf1995hitting}.

Processing-in-memory (PIM) mitigates this bottleneck by moving computation closer to where data resides~\cite{ghose2019processing,mutlu2022modern}. PIM can be categorized into two architectural forms: processing-using-memory (PUM) and processing-near-memory (PNM). PUM exploits the physical properties of non-volatile devices to execute in-situ bit-serial operations, providing high throughput for regular workloads~\cite{wong2010phase,gupta2018felix,li2016pinatubo}. Conversely, PNM integrates CMOS logic near memory banks, utilizing technologies like HBM to provide the bandwidth and flexibility required for irregular addressing logic~\cite{lee201425,loh20083d,seo2025facil}. However, most PIM accelerators are domain-specific and optimized for a single kernel, which conflicts with the compute–memory divergence in graph DP. On one end, S2G alignment is dominated by latency-sensitive, irregular memory behaviors: profiling of leading S2G software shows that both seeding and alignment suffer from high DRAM latency, driven by unpredictable pointer chasing, large intermediate tables, and high cache miss rates~\cite{cali2022segram,cali2020genasm}. On the other end, APSP workloads exhibit compute-dense min-plus updates that map well to PIM. RAPID-Graph~\cite{chen2026rapidgraph} shows strong speed and energy gains for APSP, but its datapath is too rigid for traversal-centric computation. As a result, a one-size-fits-all PIM design is underutilized across these regimes.

This compute-memory divergence defines the design of existing PIM-based graph accelerators. One class integrates fixed logic for high-performance sequence alignment, as seen in GenASM~\cite{cali2020genasm} and SeGraM~\cite{cali2022segram}, but these rigid structures cannot adapt to diverse graph DP patterns. Conversely, programmable frameworks like GenDP~\cite{gu2025gendp} offer generality at the cost of significant performance penalties compared to dedicated hardware. This inefficiency stems from the mismatch between varied dataflows and a single homogeneous architecture. These approaches face a fundamental conflict between architectural rigidity and execution efficiency. These limitations necessitate a heterogeneous solution that balances hardware specialization with general applicability.

To address this challenge, we propose GEN-Graph, a software-hardware co-designed chiplet-based heterogeneous PIM architecture for general graph-based DP. A matrix tile is designed to efficiently execute dense, matrix-like operations characteristic of algorithms such as APSP. A traversal tile is optimized for the sparse, pointer-chasing workloads found in S2G alignment. This heterogeneity optimizes resource to workload matching. Our core contributions are as follows:

\begin{itemize}
\item We perform a detailed analysis of graph-based DP algorithms and identify two fundamental and distinct computational patterns that motivate the need for a heterogeneous architecture.

\item We propose GEN-Graph, the first chiplet-based heterogeneous PIM architecture designed to efficiently accelerate both dense and sparse computational patterns.

\item We develop a compiler that implements tailored data mapping and scheduling schemes for each tile to maximize hardware utilization. This automated flow enables high efficiency across diverse graph applications.

\item We conduct a comprehensive evaluation of GEN-Graph, demonstrating a $42.8\times$ speedup over NVIDIA H100 for APSP by exploiting in-situ bit-parallelism, and a $2.56\times$ throughput gain over SOTA accelerators for S2G alignment by localizing irregular dependencies within shared banked SRAM.
\end{itemize}

\section{Background on Graph Dynamic Programming}
\label{sec:preliminary}

\begin{figure}[t]
    \centering
    \includegraphics[width=1\linewidth]{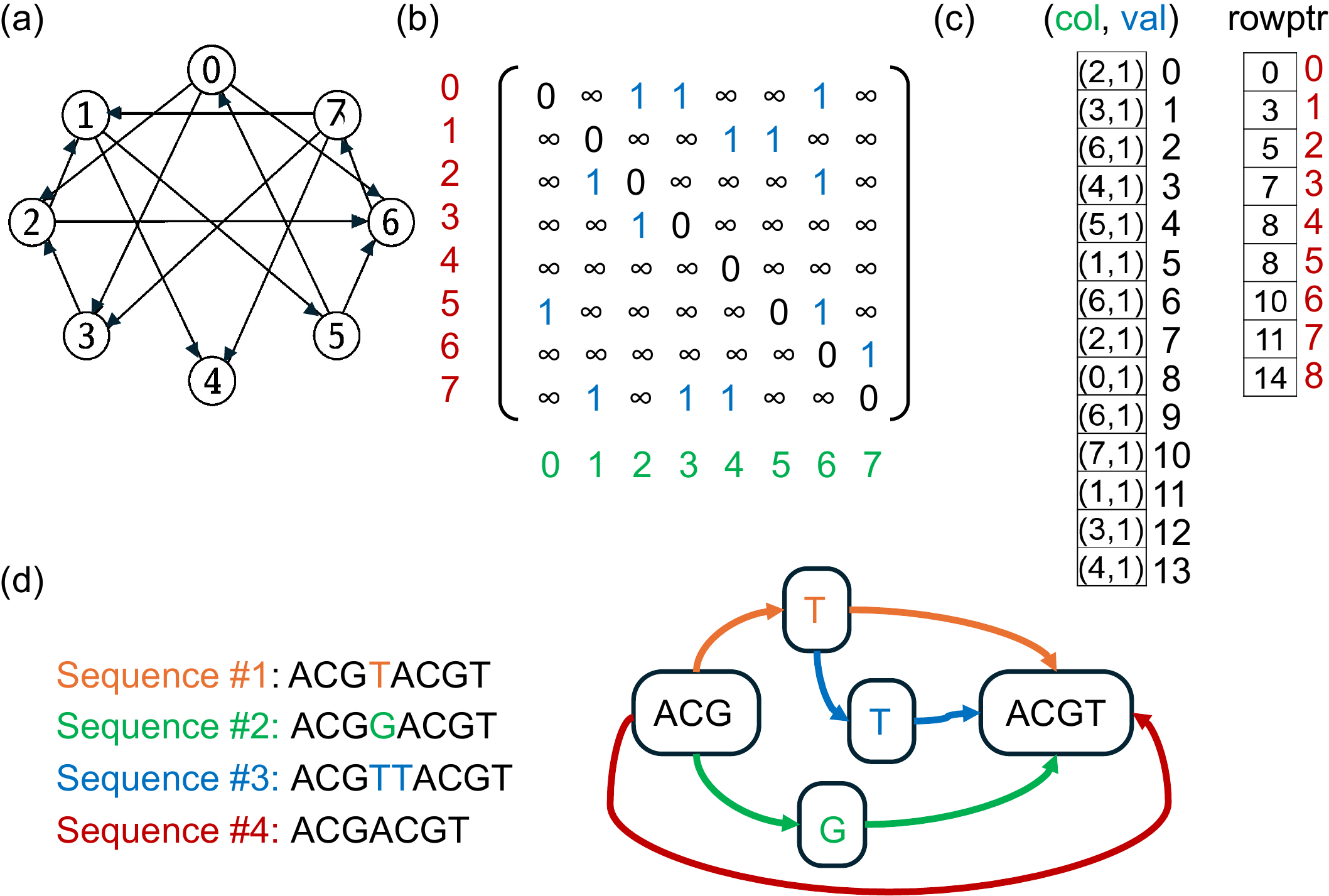}
    \caption{Graph representations (a) Topology (b) Adjacent matrix (c) Compressed sparse row (CSR) (d) Genome graph}
    \label{fig:graph}
\end{figure}

A graph $G=(V,E,w)$ is defined by its vertices $V$, edges $E$, and weights $w$. Fig.~\ref{fig:graph} illustrates an 8-vertex graph through three standard lenses: (a) the original topology, (b) a dense matrix with finite weights, and (c) the CSR format with $\{\texttt{rowptr}, \texttt{col}, \texttt{val}\}$, which optimizes storage when $|E|\ll n^2$. Beyond standard representations, genome graphs (Fig.~\ref{fig:graph}(d)) incorporate features like bubbles and joins to encode genetic diversity. These divergences define the DP archetypes described below: regular matrices enable high compute parallelism for APSP, while irregular topologies induce the sparse, non-local dependencies of sequence alignment.

\subsection{Graph-based DP for APSP} 
The classic FW algorithm~\cite{floyd1962algorithm} solves the APSP problem on a weighted graph \( G=(V,E,w) \) via an in-place DP over an \( n \times n \) distance matrix \( D \). $D[i][j]$ is set to the weight of edge $(i,j)$, or $\infty$ if no edge exists. The algorithm statically updates all entries by checking if paths through vertex $k$ yield shorter distances by:
\begin{equation}
D[i][j] = \min\left(D[i][j],\; D[i][k] + D[k][j]\right), k \in [1, n]
\end{equation}
After $n$ iterations, $D$ contains the exact shortest path lengths between all vertex pairs. The algorithm runs in \(O(n^3)\) time and \(O(n^2)\) space, following a computation pattern equal to a dense outer product of row \( i \) and column \( k \) against column \( j \). This regularity allows the workload to map effectively to systolic arrays or bit-serial compute-in-memory arrays, where performance scales linearly with logic density.

\begin{figure}[t]
    \centering
    \includegraphics[width=1\linewidth]{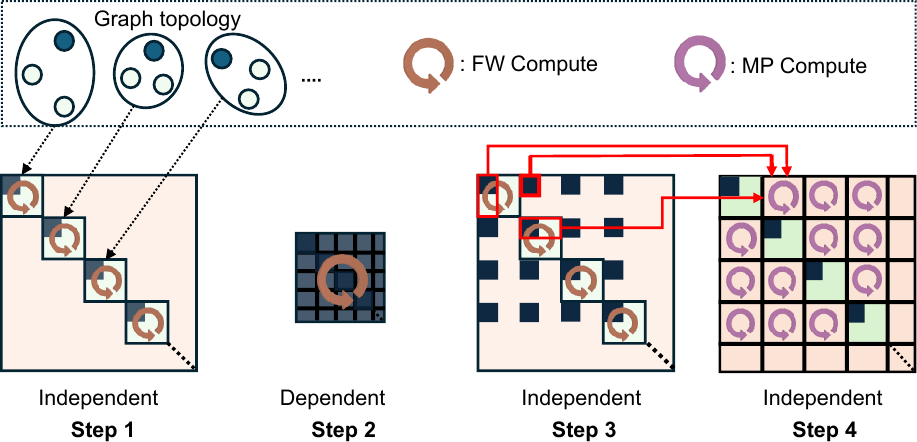}
    \caption{Illustration of partition APSP~\cite{djidjev2015all}}
    \label{fig:partitioned-apsp}
\end{figure}

The SOTA GPU algorithm partitioned APSP~\cite{djidjev2015all} is summarized in Algorithm~\ref{alg:partitioned-apsp} and illustrated in Fig.\ref{fig:partitioned-apsp}.
Graph preprocessing runs on host CPU, where a weighted graph \(G\) is partitioned into components \(C_1,\dots,C_k\) and their boundary set \(B\) via a $k$-way \textsc{Metis}~\cite{karypis1998multilevelk}. Within each component, a boundary vertex has an edge connecting to another component, while an internal vertex only has edges to vertices within its own component. The algorithm then executes in four stages:


\begin{itemize}
    
\item Step 1: Local APSP. Each component independently runs FW to fill its intra-component distance matrix \(d_\text{intra}\); all passes execute in parallel and scale linearly.

\item Step 2: Boundary-graph APSP. All boundary vertices form a reduced graph \(G_B\), with edges comprising: (i) cross-component edges from \(G\), and (ii) virtual edges within components weighted by \(d_\text{intra}\). A single FW run computes the boundary distance matrix \(d_B\). The resulting boundary distance matrix ($d_B$) involves dense $O(|B|^3)$ operations, creating the primary compute bottleneck addressed via recursive partitioning in Section~\ref{subsubsection:recursive_partitioned}.

\item Step 3: Boundary injection. Each component copies the relevant rows and columns of \(d_B\) into its local matrix and re-runs FW once, propagating inter-component shortcuts.

\item Step 4: Cross-component merge. An MP merge combines (i) source to boundary, (ii) boundary to boundary, and (iii) boundary to destination paths, producing the final cross-component distances \(d_\text{cross}\), thus completing global APSP.

\end{itemize}

\begin{algorithm}[t]
\scriptsize
\caption{Partition APSP Pseudocode}
\label{alg:partitioned-apsp}
\begin{algorithmic}[1]
\State Partition \(G\) into \(k\) components \(C_1, \dots, C_k\) via \textsc{Metis}~\cite{karypis1998multilevelk}
\State \textbf{for} \(i = 1\) \textbf{to} \(k\) \textbf{do} \hfill \Comment{Step 1}
\State \hspace{1em} \textsc{FW}\((C_i)\)
\State \(G_B \leftarrow \textsc{extractBoundaryGraph}(G)\) \hfill \Comment{Step 2}
\State \textsc{FW}\((G_B)\)
\State \textbf{for} \(i = 1\) \textbf{to} \(k\) \textbf{do} \hfill \Comment{Step 3 with injected \(d_B\)}
\State \hspace{1em} \textsc{FW}\((C_i)\)
\State \textbf{for} \(i = 1\) \textbf{to} \(k\) \textbf{do} \hfill \Comment{Step 4}
\State \hspace{1em} \textbf{for} \(j = 1\) \textbf{to} \(k\) \textbf{do}
\State \hspace{2em} \textsc{MinPlusMerge}\((C_i, C_j)\)
\State \textbf{return} global distance matrix
\end{algorithmic}
\end{algorithm}

\subsection{Graph-based DP for Genome Sequence Alignment} 

Genome graphs provide a powerful representation of genetic diversity by integrating a linear reference with known variations. Unlike a conventional linear genome that introduces reference bias toward a single individual, genome graphs represent diverse populations as a directed acyclic graph. In this model, nodes represent sequences of DNA base pairs (bps) and edges connect nodes adjacent in the genome of one or more individuals. Structural irregularities such as bubbles and joins encode specific variations, including single-nucleotide polymorphisms (SNPs) and insertions/deletions (indels)~\cite{paten2017genome}.

\begin{figure}[t]
    \centering
    \includegraphics[width=1\linewidth]{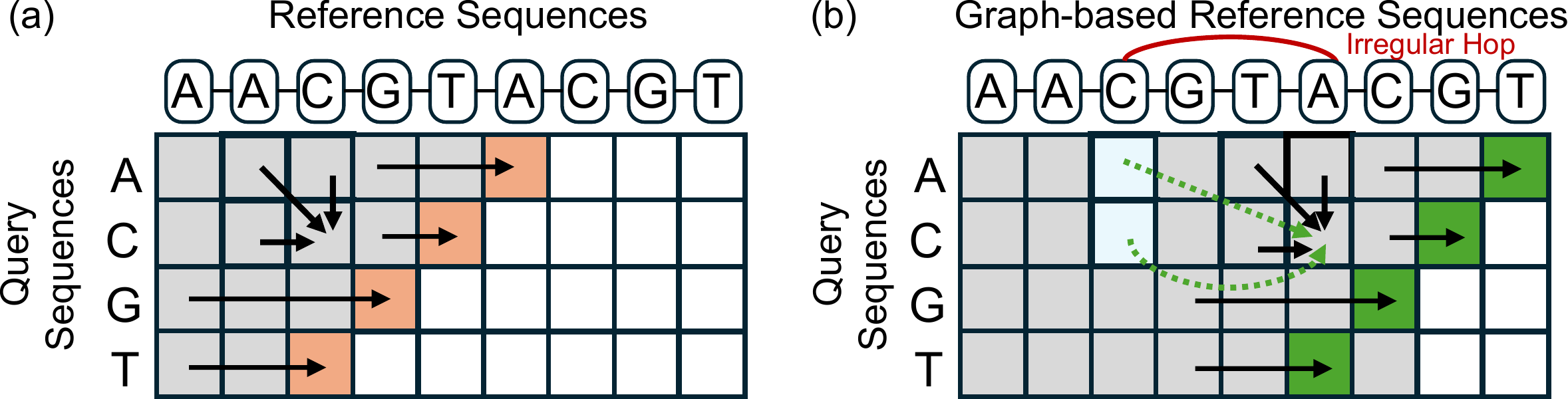}
    \caption{Genome sequencing (a) Sequence-to-sequence (b) Sequence-to-graph}
    \label{fig:genome_sequencing}
\end{figure}

In traditional sequence-to-sequence (S2S) alignment, the use of a linear reference restricts data dependencies to immediate spatial neighbors, maximizing cache locality as depicted in Fig.~\ref{fig:genome_sequencing}(a). In contrast, sequence-to-graph (S2G) alignment which incorporates genetic variations through complex graph topology as shown in Fig.~\ref{fig:genome_sequencing}(b). The S2G recurrence relation introduces topological irregularity defined by
\begin{equation}
\label{eq:s2g_recurrence}
S[v] \leftarrow \max_{u \in \text{Predecessors}(v)} (S[u] + \text{score}(u, v))
\end{equation}
where $v$ denotes the current graph vertex and $u$ represents a predecessor node. This variable set $\text{Predecessors}(v)$ creates arbitrary long-range hops. These sparse dependencies scatter memory accesses across the address space, disrupting the spatial locality and causing frequent cache misses.

Computational characteristics further bifurcate according to read length, imposing distinct processing requirements on the underlying hardware. Short-read platforms like Illumina~\cite{illumina} generate massive volumes of 100--300 bp sequences with low error rates below 1\%. The primary bottleneck here is maximizing throughput by amortizing the control overhead of millions of fine-grained, independent queries. In contrast, long-read technologies such as PacBio~\cite{pacbio_website} and Oxford Nanopore~\cite{ont_website} produce sequences ranging from 10 kbp to more than 100 kbp with higher error profiles of 5--15\%. These extensive reads create deep dependency chains within the alignment matrix, requiring sustained pipeline utilization to manage the latency of traversing large, complex graphs.

\section{Related Work}

Prior work on accelerators for graph-based DP has largely converged on domain-specific designs. Existing approaches typically optimize for either sparse traversal behavior or dense matrix-style computation, reflecting recurring computational archetypes rather than application-specific choices. Each archetype necessitates a tailored hardware strategy, a principle that has become increasingly central with the shift toward domain-specific architectures~\cite{hennessy2018new}. Accordingly, we classify existing work into three categories: topology-centric, matrix-centric, and generalized frameworks.


SeGraM~\cite{cali2022segram} integrates seeding and bit-parallel alignment into a topology-centric design that supports both S2S and S2G mapping. It achieves up to 742$\times$ speedup over software baselines and 4.8$\times$ improvement over prior accelerators. GenASM~\cite{cali2020genasm} targets approximate alignment using register-level bit vectors with minimal area and energy cost but lacks support for graph-shaped references or traceback. ASGDP~\cite{asgdp} implements the full DP in FPGA with prediction-based pruning and MaxHop control, delivering a $17.8\times$ throughput gain while preserving over $99.8\%$ alignment accuracy.


Dense workloads such as APSP benefit from regular, compute-bound execution patterns. RAPID-Graph~\cite{chen2026rapidgraph} accelerates APSP into recursively tiled MP operations on a domain-specific PCM-based accelerator and outperforms SOTA in both speed and energy. In contrast, RACE logic~\cite{madhavan2014race} encodes values as propagation delays in logic and remains effective only for regular graphs with limited control-flow irregularity.


Recent work has attempted to bridge this gap through a programmable framework. GenDP~\cite{gu2025gendp} proposes a programmable systolic array capable of supporting multiple genomics kernels, delivering 132$\times$ area-normalized speedup over CPUs. Despite its flexibility, GenDP incurs a 2.8$\times$ performance penalty compared to task-specific ASICs GenAx~\cite{fujiki2018genax} due to the inefficiency of mapping diverse computational patterns to a homogeneous substrate. This trade-off underscores the necessity of GEN-Graph's heterogeneous design, which physically decouples compute-bound and memory-bound resources.

\section{GEN-Graph: Heterogeneous PIM Design}

GEN-Graph resolves the compute-memory divergence in graph DP by mapping specialized hardware tiles to specific algorithmic dataflows. Profiling reveals orders-of-magnitude gap in arithmetic intensity that necessitates the physical decoupling of matrix-centric and topology-centric architectures (Section~\ref{sec:roofline}). We establish a memory hierarchy using HBM3, PCM, and FeNAND to align physical device strengths with varied access frequencies (Section~\ref{sec:hierarchy}).

The accelerator integrates high-density matrix tiles for cubic complexity updates and reconfigurable traversal tiles that adapt to varying sequence lengths (Sections~\ref{sec:matrix_tile} and \ref{sec:traversal_tile}). For APSP workloads,the matrix tile executes a co-designed recursive partitioned algorithm via in-situ bit-parallelism to saturate throughput during dense updates. For genomic workloads, the traversal tile implements a windowed S2G algorithm using a tiered storage hierarchy to sustain exact alignment accuracy without excessive area overhead. This co-designed accelerator enables high hardware utilization by overlapping in-situ computation with hierarchical data streaming.

\subsection{GEN-Graph Motivation}
\label{sec:roofline}
\begin{figure} [t]
    \centering
    \includegraphics[width=1\linewidth]{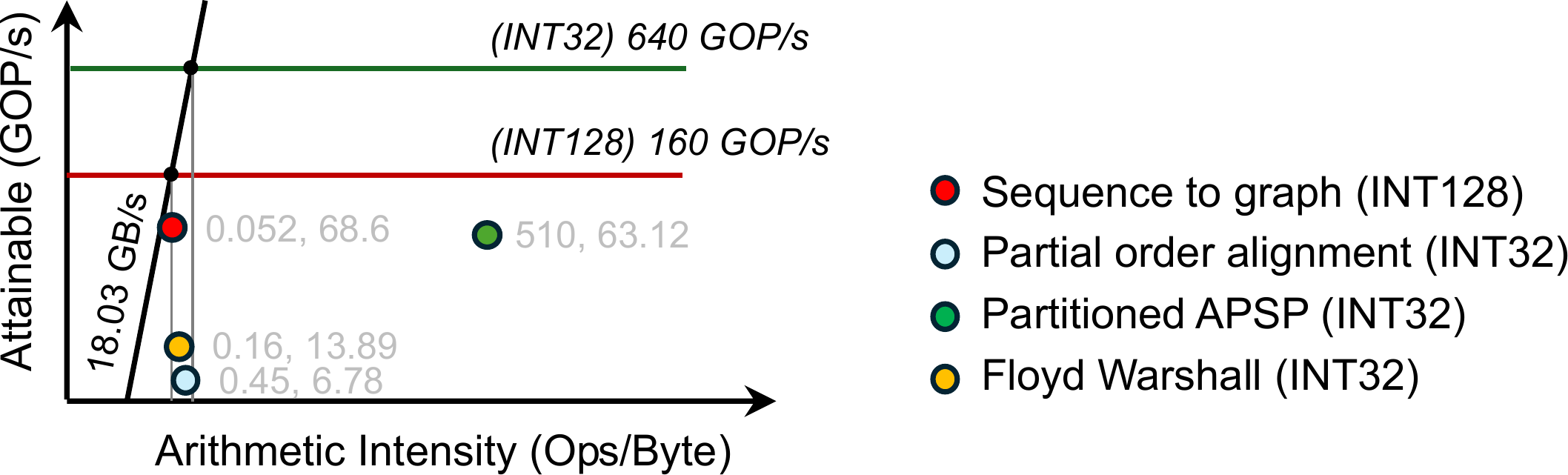}
    \caption{Roofline model of four algorithms on CPU AVX-512}
    \label{fig:roofline}
\end{figure}

Fig.~\ref{fig:roofline} provides a roofline analysis that we performed on an Intel i7-11700K (AVX-512) to quantify the compute–memory divergence between APSP and S2G. Profiling reveals orders-of-magnitude gap in arithmetic intensity between graph DP archetypes, necessitating architectural heterogeneity.

\textbf{Matrix-centric DP.} In Fig.~\ref{fig:roofline}, partitioned APSP is compute bound with an intensity of 510 Ops/Byte. It represents an optimized version of the FW algorithm (0.16 Ops/Byte). While classic FW remains memory-limited, the partitioning strategy maximizes cache residency to achieve 63.12 GOP/s. This compute-limited state motivates a PUM architecture that uses bit-serial parallelism to maximize arithmetic throughput.

\textbf{Topology-centric DP.} In Fig.~\ref{fig:roofline}, S2G and POA are memory bound, with intensities of 0.052 and 0.45 Ops/Byte respectively. Their irregular graph traversals cause frequent cache misses and force compute units to wait for data. These results confirm that data movement is the primary bottleneck for topology-based workloads. This motivates a PNM approach that integrates logic near HBM to reduce latency and improve effective bandwidth.

\subsection{Heterogeneous PIM Hierarchy}
\label{sec:hierarchy}

\begin{figure}[t]
    \centering
    \includegraphics[width=1\linewidth]{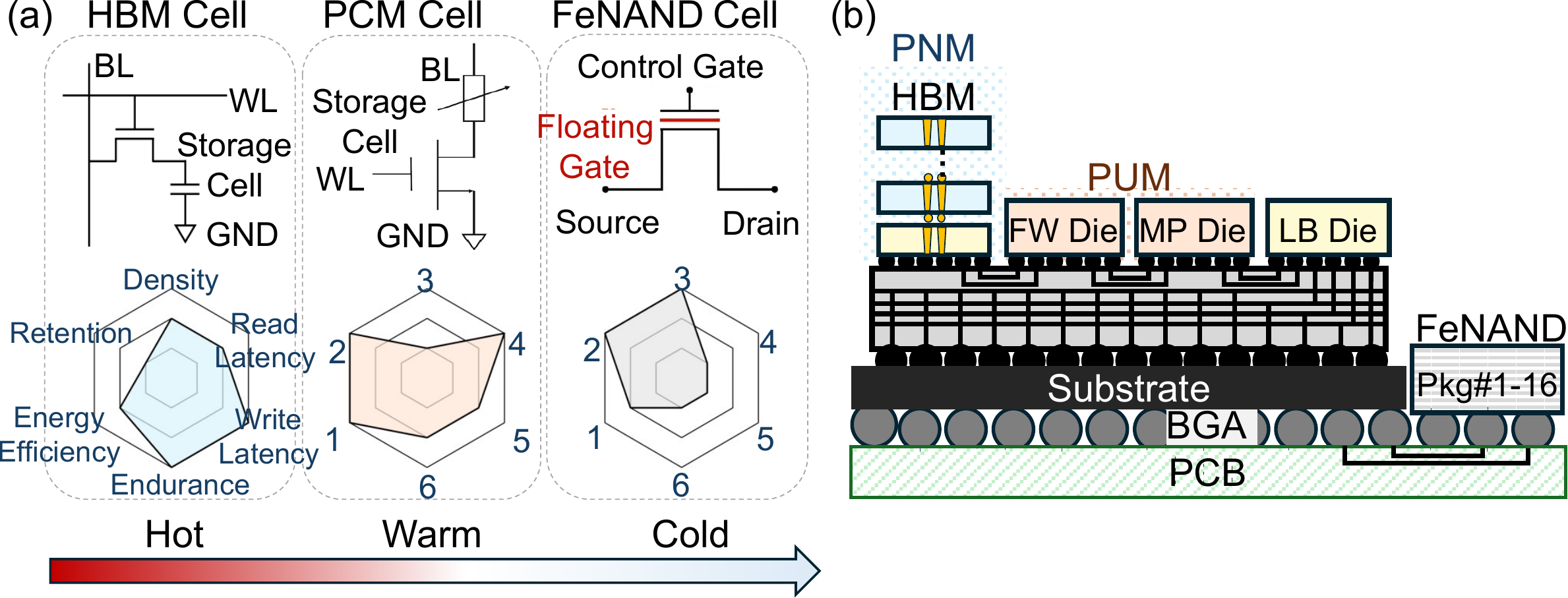}
    \caption{Heterogeneous PIM memory overview (a) Device-level tradeoffs (b) 2.5D advanced packaging cross-section}
    \label{fig:pim}
\end{figure}

The computational patterns described in motivation above highlight clear workload differences between APSP and S2G graph problems. GEN-Graph addresses this by integrating heterogeneous HBM3, PCM, and FeNAND into a complementary memory hierarchy. Fig.~\ref{fig:pim} characterizes the heterogeneous memory organization of GEN-Graph. The spider charts in Fig.~\ref{fig:pim}(a) profile the multi-dimensional trade-offs across six hardware features: density, read latency, write latency, energy efficiency, endurance, and retention. These quantitative profiles dictate a tiered memory hierarchy that aligns physical device strengths with the divergent requirements of graph DP.

\textbf{\encircle{1} Hot Tier (HBM3):} S2G alignment exhibits latency-critical irregular memory access. HBM3 dominates in write latency and write/read endurance, making it the optimal substrate for the high-frequency updates required by dynamic traversals. Programmable PNM logic on the HBM logic base die exposes this bandwidth directly to the traversal tile to minimize pipeline stalls.

\textbf{\encircle{2} Warm Tier (PCM):} APSP features compute-dense updates where energy and data persistence are critical. PCM offers superior energy efficiency and non-volatile retention. These properties enable bit-serial PUM operations without SRAM area overhead or refresh costs.

\textbf{\encircle{3} Cold Tier (FeNAND):} Pangenome graphs and large distance matrices necessitate extreme storage capacity. FeNAND excels in density and non-volatile retention, functioning as the cold tier for persistent storage. It streams static graph structures to upper tiers only when needed.

Fig.~\ref{fig:pim}(b) illustrates the physical integration via 2.5D advanced packaging. A silicon interposer connects the HBM stack and PCM dies into a unified compute package to reduce communication cost. FeNAND attaches via high-speed ONFI channels to maintain a scalable storage footprint. This organization keeps high-reuse data on-package in its most suitable compute medium.

\subsection{System Overview}
\label{sec:overview}

\begin{figure}[t]
    \centering
    \includegraphics[width=1\linewidth]{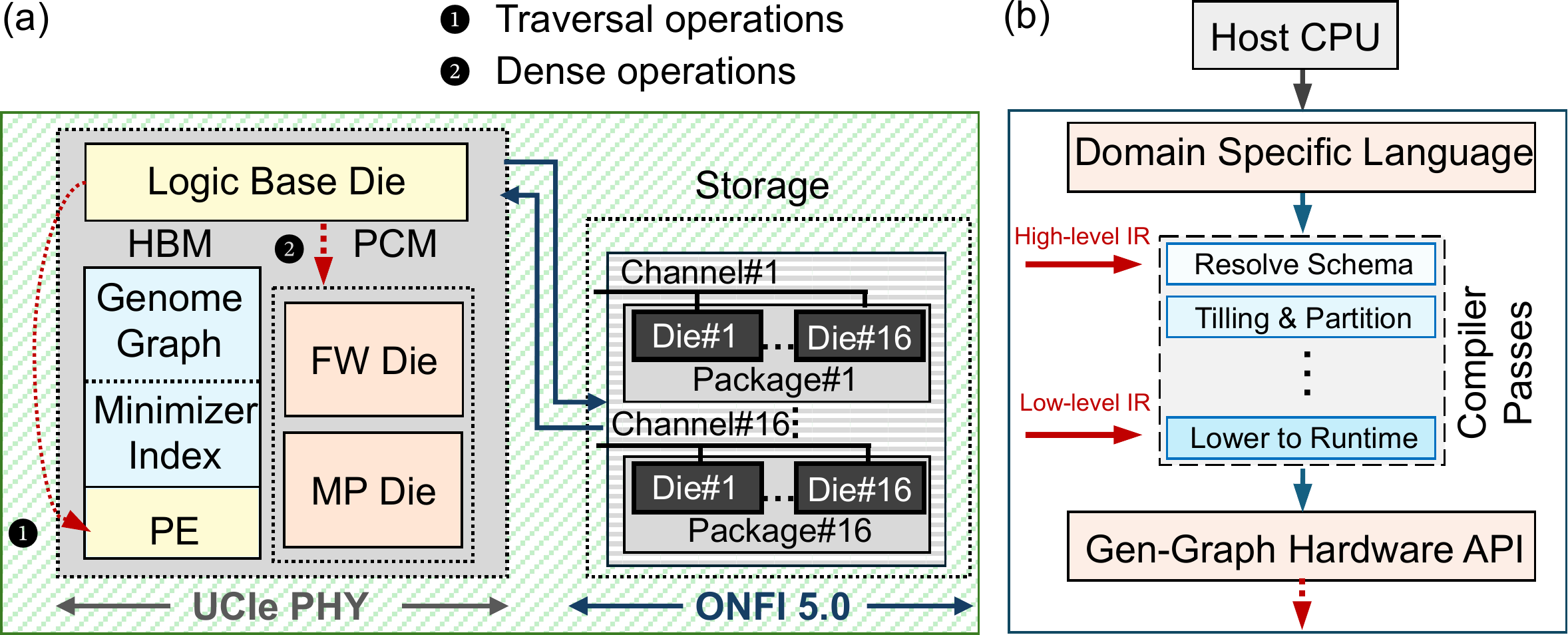}
    \caption{GEN-Graph heterogeneous PIM accelerator 
        (a) 2.5D physical architecture on the silicon interposer (b) Illustration of compilation flow}
    \label{fig:multi-die-architecture}
\end{figure}

GEN-Graph resolves the compute–memory divergence in graph DP by matching specialized compute tiles to a shared HBM–PCM–FeNAND hierarchy on a silicon interposer substrate. This co-designed accelerator enables hardware utilization by pinning data-intensive kernels to their optimal memory technology while offloading infrequent tasks to the host CPU. Fig.~\ref{fig:multi-die-architecture}(a) illustrates the 2.5D physical architecture. An HBM3 stack, PCM dies, and a logic base die integrate via a UCIe physical interface. The matrix tile serves as the PCM-based PUM engine for dense APSP updates, while the traversal tile operates as the HBM-based PNM engine for sparse S2G alignment. The logic base die integrates stream engines to orchestrate data movement across the hierarchy and off-package FeNAND via ONFI 5.1.

\begin{figure}[t]
    \centering
    \includegraphics[width=1\linewidth]{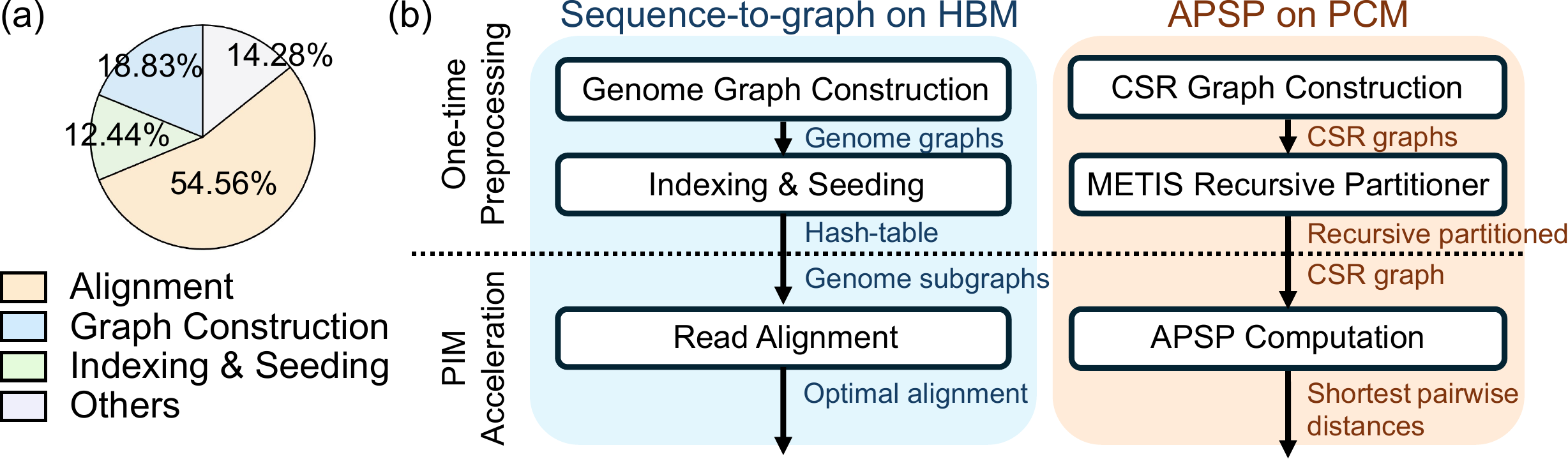}
    \caption{General graph-based DP (a) S2G runtime breakdown (b) Mapping pipeline: S2G (Left) and APSP (Right) }
    \label{fig:pie-chart}
\end{figure}

The system partitions tasks based on computational profiling to maximize execution efficiency. Based on the S2G runtime analysis in Fig.~\ref{fig:pie-chart}(a), it shows that alignment tasks consume over 70\% of total runtime, whereas host-side preprocessing overhead remains trivial. As illustrated in the mapping pipeline (Fig.~\ref{fig:pie-chart}(b)), the host CPU handles lightweight one-time preprocessing: genome graph construction, indexing, seeding and METIS~\cite{karypis1998multilevelk} partitioning. The steady state DP kernels below the dashed line instead execute on specialized PIM tiles, with S2G mapped to the traversal tile and APSP to the matrix tile. GEN-Graph therefore keeps the dominant work close to the most suitable memory technology while using the host only for light, infrequent tasks.

\begin{figure}[t]
    \centering
    \includegraphics[width=1\linewidth]{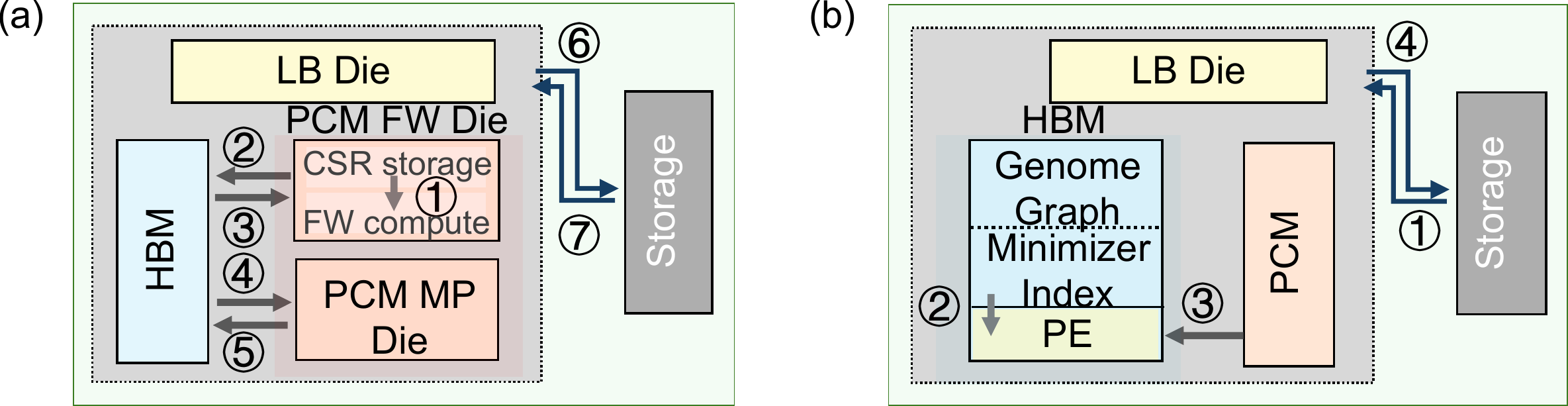}
    \caption{GEN-Graph workload dataflow 
        (a) Matrix tile (b) Traversal tile}
    \label{fig:workload-dataflow}
\end{figure}

Fig.~\ref{fig:workload-dataflow} details workload-specific dataflows. In Fig.~\ref{fig:workload-dataflow}(a) the matrix tile processes APSP. Step \encircle{1} expands CSR encoded components from HBM into dense distance tiles inside the PCM-FW die and runs bit-serial FW entirely in place. Step \encircle{2} writes updated tiles back to HBM. Step \encircle{3} extracts boundary rows and columns on the logic base die and assembles the boundary graph. Step \encircle{4} streams boundary tiles into the PCM-MP die, which performs MP reductions on boundary-to-boundary paths. Step \encircle{5} returns the reduced boundary tiles to HBM and injects them back into component tiles for the next FW phase. After a recursion level completes, Step \encircle{6} compresses dense tiles back to CSR and Step \encircle{7} streams them to FeNAND. This streaming schedule overlaps PCM computation with HBM-side boundary construction and FeNAND I/O, enabling APSP to operate at high sustained throughput on the matrix tile. In Fig.~\ref{fig:workload-dataflow}(b) the traversal tile executes S2G alignment. The full pangenome graph and minimizer index reside in FeNAND. During initialization, Step \encircle{1} streams these structures through the logic base die, builds a working copy in HBM and pins highly reused graph fragments and index buckets in PCM as a non-volatile cache. For each batch of reads, Step \encircle{2} runs the bit vector alignment kernel on PEs attached to HBM channels, using subgraphs stored in HBM and the precomputed seeds. When a batch needs cached subgraphs, Step \encircle{3} fetches the corresponding subgraph from PCM into HBM and forwards it to the PEs so hotspots bypass FeNAND. After a batch completes, Step \encircle{4} writes compact alignment results or intermediate states from HBM back to FeNAND when long term storage is required.

Finally, a domain-specific compiler drives the hardware (Fig.~\ref{fig:multi-die-architecture}(b)) by translating domain-specific language (DSL) on the host into accelerator commands. The compiler first translates the program into a high-level intermediate representation (IR), then resolves schemas and applies key transformations such as tiling and recursive partitioning to match the hardware tiles. Next, it lowers the optimized IR into a runtime-level representation and emits GEN-Graph hardware API calls that drive execution on the heterogeneous PIM accelerator. The generated runtime calls select the appropriate tile, allocate buffers in HBM or PCM, and orchestrate data transfers between FeNAND and compute tiles to optimize data locality.

\subsection{Matrix Tile: APSP Optimized Compute Tile}
\label{sec:matrix_tile}

The matrix tile scales APSP by partitioning large graphs into 1024-vertex clusters that match physical PIM capacities (Section~\ref{subsubsection:recursive_partitioned}). In Section~\ref{subsubsection:pcm-die-arch}, we describe the hardware architecture and specialized logic units where PCM-FW dies execute data permutations and PCM-MP dies accelerate min-reductions. Finally, tailored mapping and scheduling schemes drive bit-serial execution to saturate near-memory bandwidth and maximize parallel throughput (Section~\ref{subsubsection:mapping}).

\subsubsection{Co-Designed Recursive Partitioned APSP}
\label{subsubsection:recursive_partitioned}

\begin{figure}[t]
    \centering
    \includegraphics[width=1\linewidth]{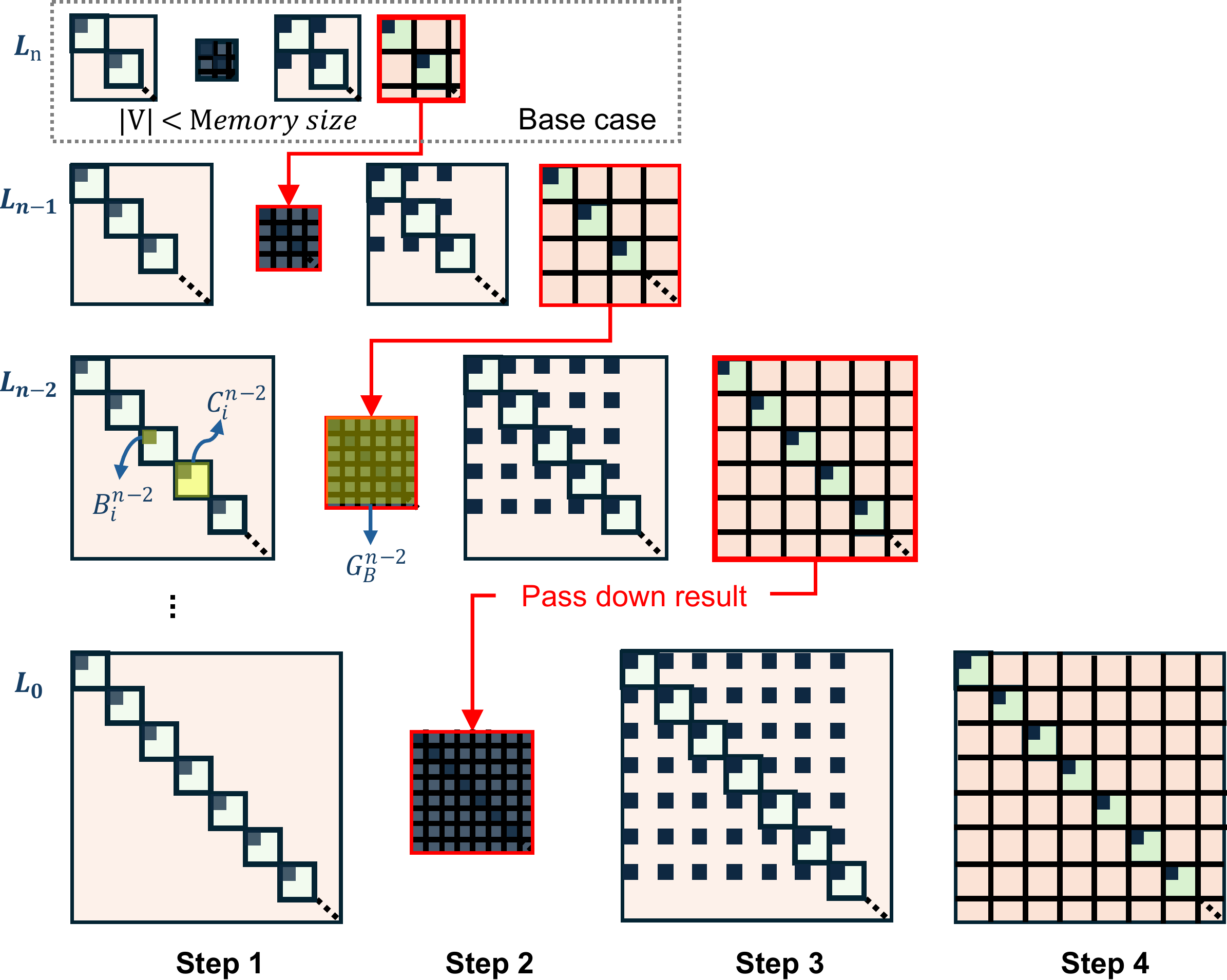}
    \caption{Illustration of Recursive Partition APSP}
    \label{fig:recursion-apsp-flow}
\end{figure}

To efficiently scale APSP computation on large graphs, we design a recursive partitioning strategy that enables fully independent subgraphs sized to fit within PIM tile limits. Algorithm~\ref{alg:recursive-apsp} summarizes the recursive APSP procedure across hierarchy levels. It follows the same four steps as Algorithm~\ref{alg:partitioned-apsp}, but operates bottom-up: starting from base-level partitions, each level computes local APSP and propagates boundary summaries upward. Accordingly, we partition each component at $|V| \leq 1024$, matching practical array dimensions per PCM tile and the maximum parallelism achievable with dense, high-yield fabrication. The input graph $G=(V,E,w)$, with vertex set $V$, edge set $E$, and non-negative weights $w$, is first partitioned by METIS~\cite{karypis1998multilevelk} into base-level components $C_1^{(0)}, \dots, C_k^{(0)}$. Each component $C_i^{(0)}$ contains internal vertices and boundary vertices, where boundary vertices connect to other components. We extract boundary vertices from ${C_i^{(0)}}$ to construct the level-0 boundary graph $G_B^{(0)}$. If boundary graph $G_B^{(\ell)}$ at level $\ell$ exceeds the 1024-vertex tile limit, we recursively partition it, creating a coarser graph $G_B^{(\ell+1)}$. This continues until $|V(G_B^{(n)})|\leq1024$, ensuring all graphs fit entirely within tiles. Table~\ref{tab:variables} summarizes key variables. At each recursion level, APSP is computed locally within components and boundary graphs, propagating distances back into components and performing cross-component updates via MP products. Fig.~\ref{fig:recursion-apsp-flow} illustrates this process. After recursive partitioning, each memory array holds one dense distance block with \(N \le 1024\) vertices. To maximize parallelism during the FW updates, we adopt a specialized data remapping strategy. This strategy logically separates the current pivot row and column (the \texttt{Panel\_Row} and \texttt{Panel\_Col}) from the rest of the distance matrix (the \texttt{Main\_Block}). This structure maximizes tile-level parallelism on the \texttt{Main\_Block} and fits within constrained PIM resources without requiring global synchronization. The detailed mapping and scheduling schemes for this process are described in Section~\ref{subsubsection:mapping}.

\begin{table}[t]
\caption{Key Variables in Recursive Partitioned APSP}
\label{tab:variables}
\begingroup
\renewcommand{\arraystretch}{1.25}
\centering
\scalebox{1}{
\begin{tabular}{|l|l|}
\hline
\multicolumn{1}{|c|}{\textbf{Variable}} &
\multicolumn{1}{c|}{\textbf{Description}} \\
\hline
$G = (V,E,w)$ & Graph with vertices $V$, edges $E$, edge weights $w$ \\
$C_i^{(\ell)}$ & Component at level $\ell$ ($\ell$=1,...,n and i=1,...,k) \\
$B_i^{(\ell)}$ & Boundary vertex set of component $C_i^{(\ell)}$ \\
$G_B^{(\ell)}$ & Level-$\ell$ boundary graphs \\
$\mathrm{DB}^{(\ell)}$ & Boundary-to-boundary distance matrix at level $\ell$ \\
$D_C$ & Intra-component APSP distance of component $C$ \\
$D_{C_1}[m,n]$ & Cross-component distance from $m\in C_1$ to $n\in C_2$\\
\hline
\end{tabular}
}
\endgroup
\end{table}

\begin{algorithm}[t]
\scriptsize
\caption{Recursive Partition APSP Pseudocode}
\label{alg:recursive-apsp}
\begin{algorithmic}[1]
\State \textbf{for} \(\ell = n\) down to \(0\) \textbf{do}
\State \hspace{1em} \textbf{parallel for} \(C\) in levels[\(\ell\)] \textbf{do} \hfill \Comment{Step 1}
\State \hspace{2em} \(D_C \gets \text{FloydWarshall}(C)\)
\State \hspace{2em} \(B_C \gets \text{find\_boundary}(C)\)
\State \hspace{2em} \textbf{if} DB\_prev == \(\varnothing\): \(DB_C \gets \text{restrict}(D_C, B_C)\)
\State \hspace{1em} \textbf{if} DB\_prev == \(\varnothing\): \hfill \Comment{Step 2}
\State \hspace{2em} \(G_B \gets \text{build\_boundary\_graph}(\{DB_C \text{ for all } C\})\)
\State \hspace{2em} DB\_prev \(\gets \text{FloydWarshall}(G_B)\)
\State \hspace{1em} \textbf{parallel for} \(C\) in levels[\(\ell\)] \textbf{do} \hfill \Comment{Step 3}
\State \hspace{2em} \(D_C \gets \text{inject}(DB\_prev, B_C)\)
\State \hspace{1em} \textbf{parallel for} \((C_1, C_2)\) in levels[\(\ell\)] \textbf{do} \hfill \Comment{Step 4}
\State \hspace{2em} \textbf{for} \(m\) in \(B_{C_1}\), \(n\) in \(B_{C_2}\) \textbf{do}
\State \hspace{2em} \(D_{C_1}[m,n] \gets \min\limits_{\substack{i \in B_{C_1} \\ j \in B_{C_2}}} \left( D_{C_1}[m,i] + \text{DB\_prev}[i,j] + D_{C_2}[j,n] \right)\)
\State \hspace{1em} DB\_prev \(\gets \text{merge}(\{\text{restrict}(D_C, B_C)\})\) 
\State \textbf{return} DB\_prev
\end{algorithmic}
\end{algorithm}

\begin{figure}[t]
    \centering
    \includegraphics[width=1\linewidth]{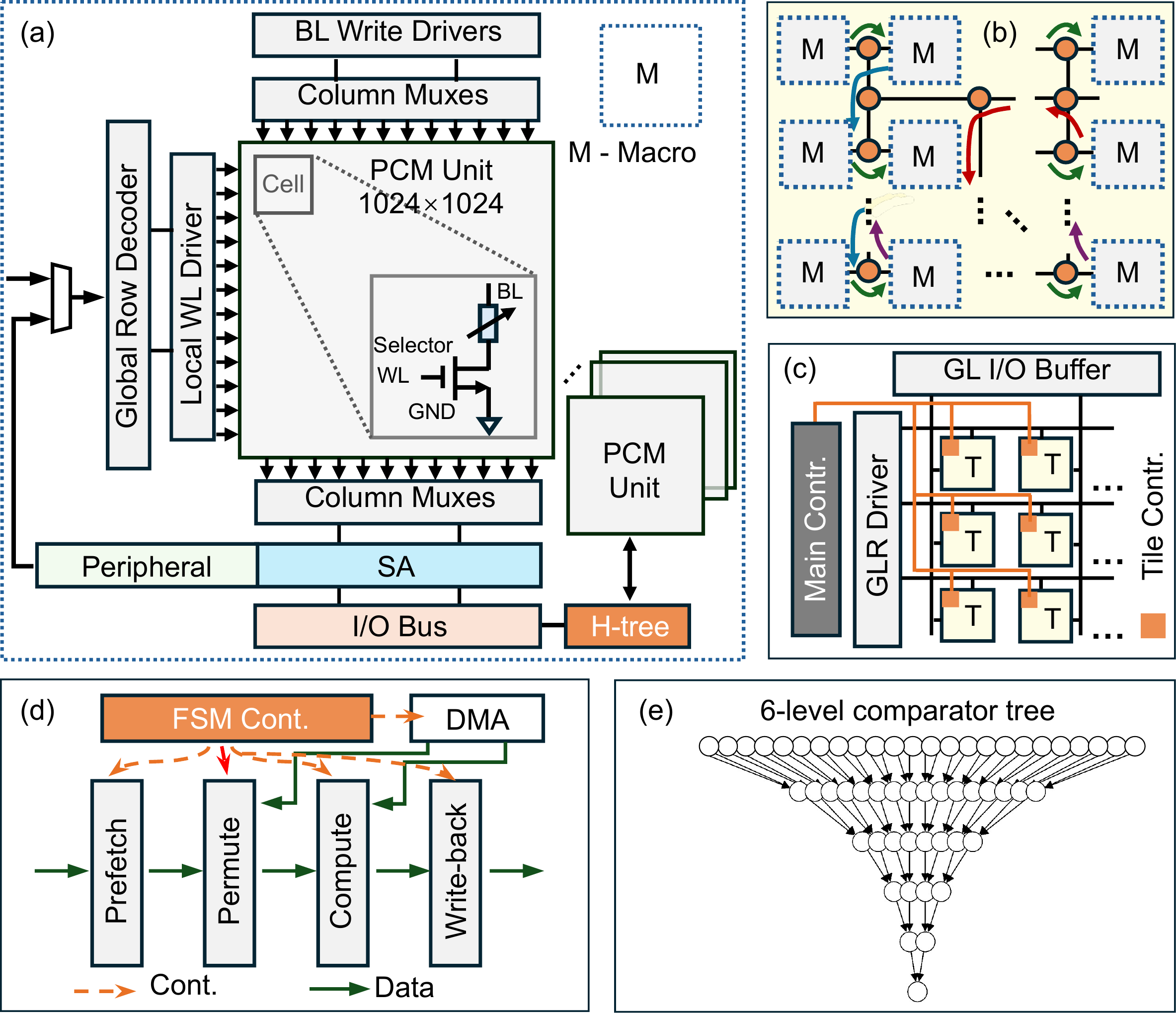}
    \caption{Matrix tile detail architecture (a) Macro (M) (b) Tile (T) (c) Die (d) Permutation unit (e) 6-level min comparator tree}
    \label{fig:pim_arch}
\end{figure}

\subsubsection{Matrix Tile Hardware Design and Implementation}
\label{subsubsection:pcm-die-arch}
Both PCM dies use the same 1T1R SLC PCM cell technology~\cite{wu2024novel}, organized into $1024 \times 1024$ units. Each unit, together with its peripheral circuits, forms a macro, as shown in Fig.\ref{fig:pim_arch}(a). Each tile contains 130 parallel units connected via an H-tree interconnect~\cite{7551379} for efficient data exchange (Fig.~\ref{fig:pim_arch}(b)). The design ensures full crossbar activity without idle cycles. Matching arrow colors denote concurrent unit-to-unit transfers.
Tiles operate concurrently, each with a local controller linked to a global main controller for coordinated execution (Fig.~\ref{fig:pim_arch}(c)). In-memory operations are triggered by row-segment broadcasts, with each tile controller managing parallel processing across its units. In addition to shared peripheral logic, each PCM die integrates specialized circuits for its role. The PCM-FW tile includes a permutation unit for rearranging data blocks, while the PCM-MP tile integrates a 32-bit 6-level min-comparator tree for efficient MP reductions.

\textbf{PCM-FW Permutation Unit.} 
The PCM-FW die includes a dedicated permutation macro that locally rearranges data blocks, avoiding off-die data movement. As shown in Fig.~\ref{fig:pim_arch}(d), the permutation macro comprises:

\begin{enumerate}
  \item a 32-row burst row-buffer controller,
  \item a reorder buffer for panel masking and block pruning,
  \item a lightweight on-tile DMA engine (1-cycle read, 10-cycle write) with address remapper, and
  \item a four-stage FSM pipeline \small(Prefetch $\!\rightarrow$ Permute $\!\rightarrow$ Compute $\!\rightarrow$ Write-back\small),
\end{enumerate}
so that data movement overlaps computation. The permutation unit packs \texttt{Panel\_Row} and mirrored \texttt{Panel\_Col} into 32-row windows for coalesced bursts without H-tree stalls, while a \texttt{prefetch} buffer hides DMA latency and skips futile writes, sustaining near-peak occupancy and reducing wear.

\textbf{PCM-MP Min-Compare \& Update.} 
Each unit integrates a pipelined comparator tree (Fig.~\ref{fig:pim_arch}(e)) that reduces 1024 32-bit inputs to one 32-bit minimum in following steps: A 1024 × 32-bit row is streamed into the buffer in 1 cycle. Thirty-two parallel five-level carry look-ahead (CLA) trees extract sign bits and 5-bit indices for block minima in 6 cycles. A second five-level tree reduces these to a global minimum in another 6 cycles, totaling 13 cycles per row. The final sign-bit mask gates PCM writes, updating only entries with smaller values. Two staging buffers hold operands across adds while $DB[i,j]$ streams, enabling one 1024-wide vector per cycle. The reduction tree outputs an update mask that enables compare-and-swap selective writes, avoiding read–modify–write and lowering energy and wear.

\begin{figure}[t]
    \centering
    \includegraphics[width=1\linewidth]{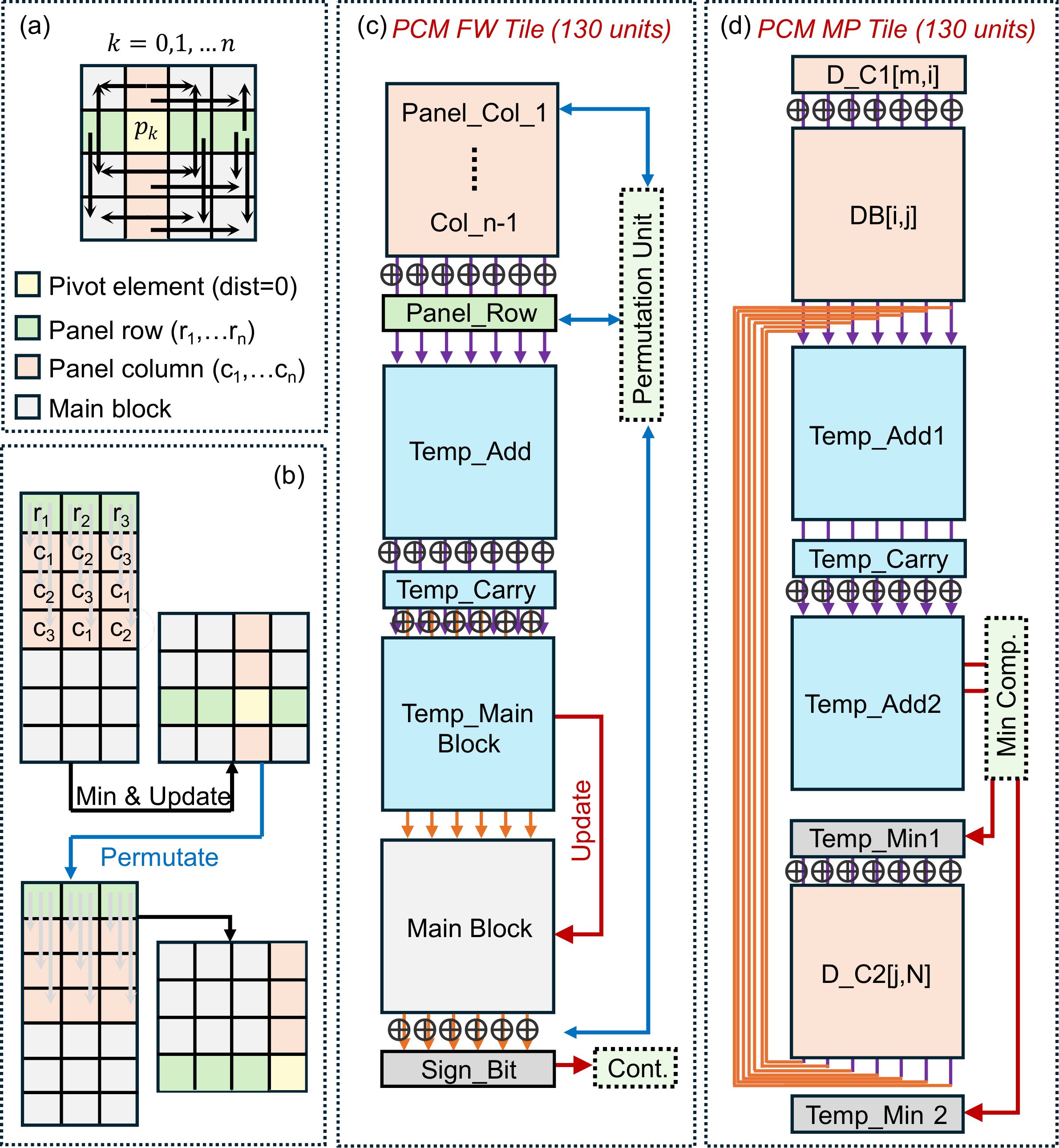}
    \caption{SW/HW mapping for recursive partitioned APSP (a) FW illustration (b) FW remapping, every step includes add, min update and permutation (c) PCM-FW tile performing FW (d) PCM-MP tile performing two-stage MP}
    \label{fig:Mapping}
\end{figure}

\subsubsection{APSP Mapping and Scheduling Scheme}
\label{subsubsection:mapping}

To execute FW and MP efficiently in PCM tiles, we co-optimize software layout and hardware design through tailored intra-tile mapping and scheduling. Fig.~\ref{fig:Mapping}(a) illustrates the FW DP update flow. Since diagonal pivot elements $p_k$ always have zero distance, their propagation along pivot row ($r_1$,...$r_n$) and column ($c_1$,...$c_n$) is omitted. Instead, row and column elements propagate into the main block to perform add and min operations, updating main block values with MP products. Each pivot element triggers one such update; the process repeats until all diagonal pivots $p_k$ ($k$=1,..,n) are processed. Fig.~\ref{fig:Mapping}(b) shows our remapping strategy that maximizes array parallelism. We extract the pivot row, copy its column beneath it, and form an $(n{-}1) \times (n{-}1)$ block; this lets all updates finish with one add and one min. The next pivot element is then permuted, and the process repeats.

Both PCM-FW and PCM-MP tiles contain 130 units and are partitioned into regular functional regions. In the PCM-FW tile, the pivot column is stored in \textcolor[RGB]{180,130,100}{\texttt{Panel\_Col}[1:\,$n-1$]}, the pivot row is stored in \textcolor[RGB]{130,180,130}{\texttt{Panel\_Row}}, and the remaining entries reside in \textcolor[RGB]{110,110,110}{\texttt{Main\_Block}}. Intermediate values are buffered in \textcolor[RGB]{80,150,180}{\texttt{Temp\_Add}}, \textcolor[RGB]{80,150,180}{\texttt{Temp\_Carry}}, and \textcolor[RGB]{80,150,180}{\texttt{Temp\_Main\_Block}}, while the final sign bit is recorded in \textcolor[RGB]{40,40,40}{\texttt{Sign\_Bit}}. Fig.~\ref{fig:Mapping}(c) shows the PCM-FW datapath. Add is implemented with FELIX~\cite{gupta2018felix} bit-serial adders using \textcolor[RGB]{80,150,180}{\texttt{Temp\_Add}} and \textcolor[RGB]{80,150,180}{\texttt{Temp\_Carry}}. A FELIX~\cite{gupta2018felix} bit-serial subtraction compares \textcolor[RGB]{80,150,180}{\texttt{Temp\_Main\_Block}} against \textcolor[RGB]{110,110,110}{\texttt{Main\_Block}}, and the resulting sign bit gates selective writes back to \textcolor[RGB]{110,110,110}{\texttt{Main\_Block}}. The permutation unit then reorders \textcolor[RGB]{130,180,130}{\texttt{Panel\_Row}}, \textcolor[RGB]{180,130,100}{\texttt{Panel\_Col}}, and \textcolor[RGB]{110,110,110}{\texttt{Main\_Block}} for the next pivot, as detailed in Section~\ref{subsubsection:pcm-die-arch}. Fig.~\ref{fig:Mapping}(d) shows the PCM-MP tile, which also comprises 130 units and performs a two-stage MP merge. In the first stage, a logical row $(1 \times 1024)$ from $C_1$ is read into \textcolor[RGB]{180,130,100}{\ensuremath{D_{C_1}[m,i]}}, while \textcolor[RGB]{180,130,100}{\ensuremath{DB[i,j]}} and the corresponding row \textcolor[RGB]{180,130,100}{\ensuremath{D_{C_2}[j,n]}} are presented in parallel. The merge applies two successive MP steps, \textcolor[RGB]{180,130,100}{\ensuremath{D_{C_1}[m,i]}} + \textcolor[RGB]{180,130,100}{\ensuremath{DB[i,j]}} followed by $(\cdot)$ + \textcolor[RGB]{180,130,100}{\ensuremath{D_{C_2}[j,n]}}, each followed by a 1024-way min reduction. Computation uses FELIX bit-serial adders in \textcolor[RGB]{80,150,180}{\texttt{Temp\_Add1}}, \textcolor[RGB]{80,150,180}{\texttt{Temp\_Carry}}, and \textcolor[RGB]{80,150,180}{\texttt{Temp\_Add2}}, and a pipelined comparator tree reduces all candidates to a single minimum stored in \textcolor[RGB]{100,100,100}{\texttt{Temp\_Min1}} and \textcolor[RGB]{100,100,100}{\texttt{Temp\_Min2}}.

\subsection{Traversal Tile: S2G Optimized Compute Tile}
\label{sec:traversal_tile}

The traversal tile co-designs the windowed bit-parallel algorithm to resolve the complex topological dependencies (Section~\ref{subsubsection:s2g}). To mitigate memory bottlenecks, the hardware integrates a tiered storage hierarchy on HBM3 logic dies to stage alignment states and pattern masks (Section~\ref{subsubsection:s2ghw}). This organization sustains high utilization via an adaptive mapping scheme that reconfigures processing elements (PEs) for high-throughput parallel execution of short reads or deep pipelining for long sequences (Section~\ref{subsubsection:s2gmapping}).

\subsubsection{Sequence-to-Graph Alignment}
\label{subsubsection:s2g}
The traversal tile implements S2G alignment using a windowed bit-parallel DP formulation. It represents the DP wavefront using dense bit vectors and expresses state updates as simple bitwise operations such as AND, OR, and shifts.

As detailed in Algorithm~\ref{alg:s2g_logic_clean}, the process begins by partitioning the query $Q$ into $k$ segments $Q_i$ to match the hardware vector width $W$. For each segment, precomputed match masks $\mathcal{M}$ are generated for the current window. The inner loop processes graph nodes $v$ in topological order to resolve two-dimensional dependencies. First, spatial dependencies are resolved by aggregating predecessor state vectors $\mathcal{S}[u]$ into $\vec{D}_{in}$ via bitwise OR ($\bigvee$) operations. Second, temporal dependencies across windows are resolved by computing the updated state $\vec{D}_{v}$ through a bitwise shift of $\vec{D}_{in}$ and the injection of a carry-in bit $c_{in}$. This $c_{in}$ is initialized to 1 for the first segment and retrieves the carry bit $C[v]$ from the previous segment otherwise. Finally, the most significant bit (MSB) of $\vec{D}_{v}$ is stored in $C[v]$ for the next segment iteration, while $\mathcal{S}[v]$ is updated to record the alignment state and track the global $Score_{max}$. Windowed bit-parallelism reduces the computational complexity from $O(N \cdot M)$ to $O(N \cdot M / W)$, providing the throughput necessary for pangenome-scale alignment. Table~\ref{tab:s2g_vars} summarizes the key variables. By decoupling control flow from data movement, this organization enables parallel window execution.

\begin{table}[t]
\caption{Key Variables in Windowed S2G Alignment}
\label{tab:s2g_vars}
\begingroup
\renewcommand{\arraystretch}{1.2}
\centering
\begin{tabular}{|l|p{0.7\linewidth}|}
\hline
\textbf{Variable} & \multicolumn{1}{c|}{\textbf{Description}} \\ \hline
$Q, W$ & Query sequence and vector width ($W=128$ bits) \\
$k$ & Number of window segments ($k = \lceil |Q|/W \rceil$) \\
$Q_i$ & The $i$-th query segment ($1 \le i \le k$) \\
$\mathcal{M}$ & Precomputed match bit-masks for segment $Q_i$ \\
$\mathcal{S}[v]$ & Alignment state bit-vector for graph node $v$ \\
$\vec{D}_{in}$ & Aggregated predecessor state (Spatial Dependency) \\
$C[v]$ & Carry bit for node $v$ (Temporal Dependency) \\
$Score_{max}$ & The global maximum alignment score recorded \\ \hline
\end{tabular}
\endgroup
\end{table}

\begin{algorithm}[t]
\scriptsize
\caption{Windowed Bit-Parallel S2G Alignment}
\label{alg:s2g_logic_clean}
\begin{algorithmic}[1]
\State \textbf{Input:} Graph $G(V, E)$, Query $Q$, Window Width $W$
\State \textbf{Output:} $Score_{max}$
\Statex
\State $k \gets \lceil |Q| / W \rceil$; \quad Initialize $\mathcal{S}[v] \gets \vec{0}, C[v] \gets 1 \ \forall v \in V$

\State \textbf{for} $i \gets 1$ \textbf{to} $k$ \textbf{do} \hfill \Comment{Outer loop: Query segments}
    \State \hspace{1em} $\mathcal{M} \gets \text{PrecomputeMasks}(Q_i)$ 
    \State \hspace{1em} \textbf{for} \textbf{each} $v \in V$ \textbf{in Topological Order} \textbf{do}
        \State \hspace{2em} \textbf{Step 1: Spatial Dependency Aggregation}
        \State \hspace{2em} $\vec{D}_{in} \gets \bigvee_{u \in \text{Pred}(v)} \mathcal{S}[u]$ \hfill \Comment{Resolve graph topology}
        
        \State \hspace{2em} \textbf{Step 2: Bit-Parallel State Update}
        \State \hspace{2em} $c_{in} \gets (i == 1) ? 1 : C[v]$ \hfill \Comment{Inject inter-window carry}
        \State \hspace{2em} $\vec{S}_{new} \gets ((\vec{D}_{in} \ll 1) \lor c_{in}) \land \mathcal{M}[\text{char}(v)]$
        
        \State \hspace{2em} \textbf{Step 3: State and Carry Preservation}
        \State \hspace{2em} $\mathcal{S}[v] \gets \vec{S}_{new}$ 
        \State \hspace{2em} $C[v] \gets \text{GetMSB}(\vec{S}_{new})$ \hfill \Comment{Store for next window}
        
        \State \hspace{2em} \textbf{Step 4: Score Logging}
        \State \hspace{2em} \textbf{if} $\vec{S}_{new} \neq \vec{0}$ \textbf{then} Update $Score_{max}$ 

\State \textbf{return} $Score_{max}$
\end{algorithmic}
\end{algorithm}

\subsubsection{Traversal Tile Hardware Design and Implementation}
\label{subsubsection:s2ghw}

\begin{figure}[t]
    \centering
    \includegraphics[width=1\linewidth]{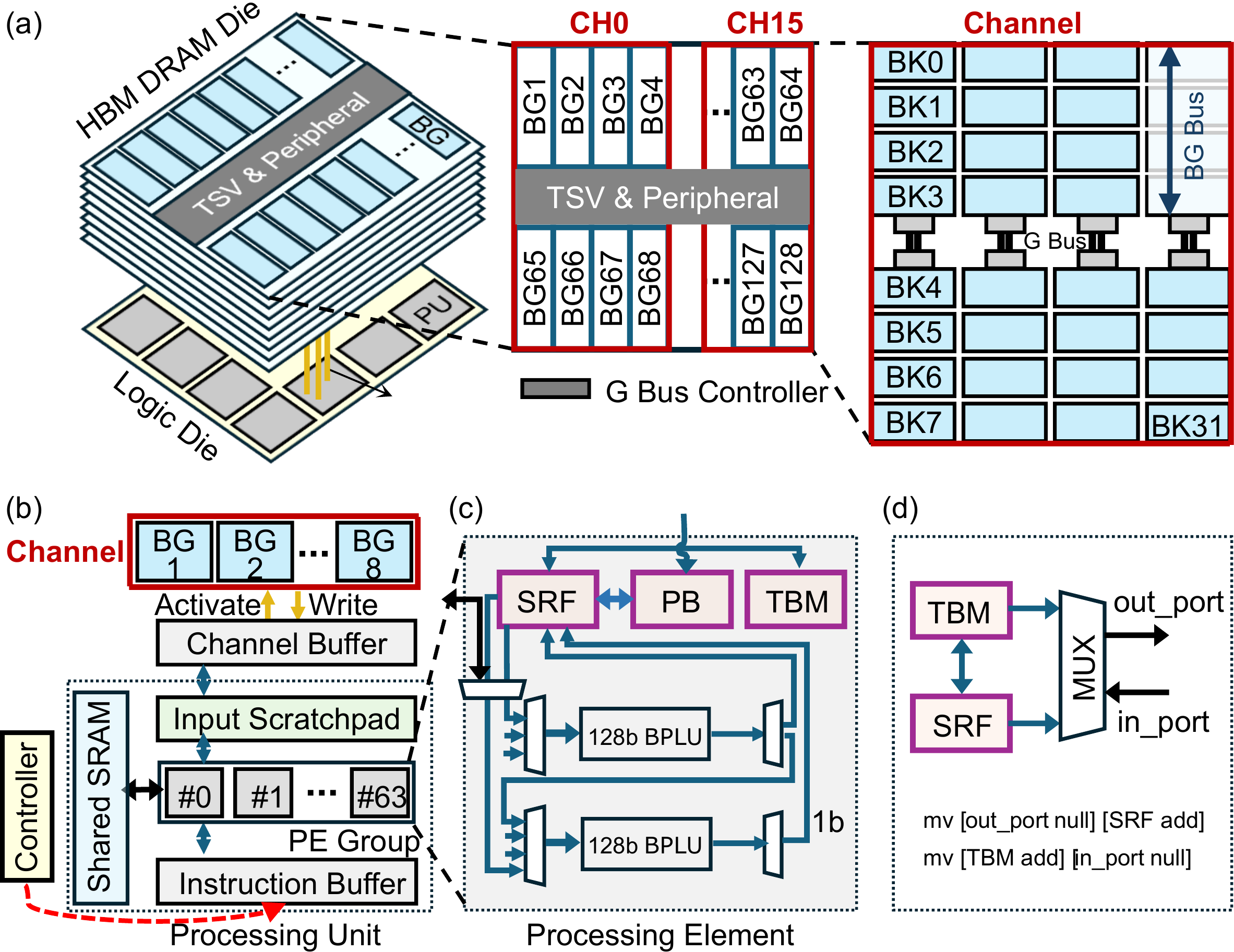}
    \caption{Traversal tile detail architecture (a) HBM3 organization (b) Processing unit (c) Processing element (d) PE data movement} 
    \label{fig:HBM-PNM}
\end{figure}

\begin{figure}[t]
    \centering
    \includegraphics[width=1\linewidth]{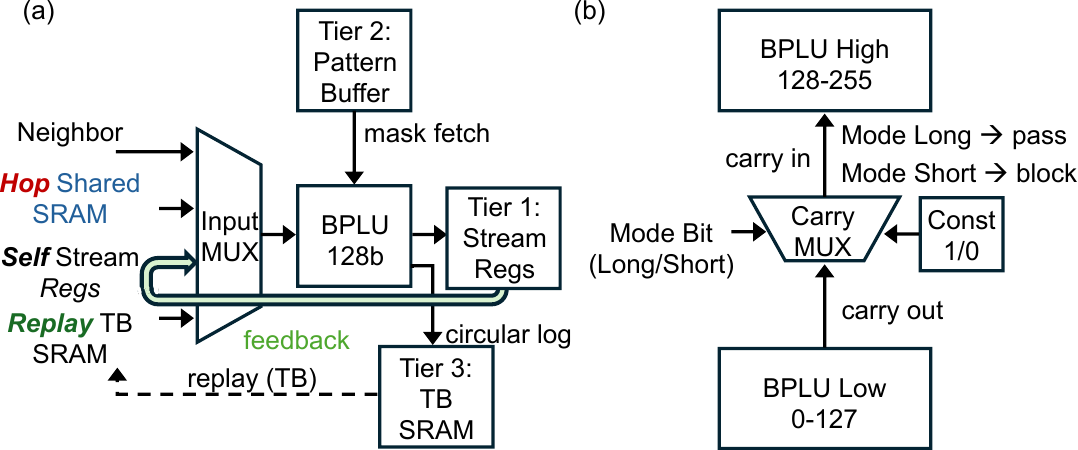}
    \caption{Refined HBM3 PNM micro architecture design (a) Three-tier storage hierarchy (b) Configurable carry chain}
    \label{fig:micro-arch}
\end{figure}

\begin{figure}[t]
    \centering
    \includegraphics[width=1\linewidth]{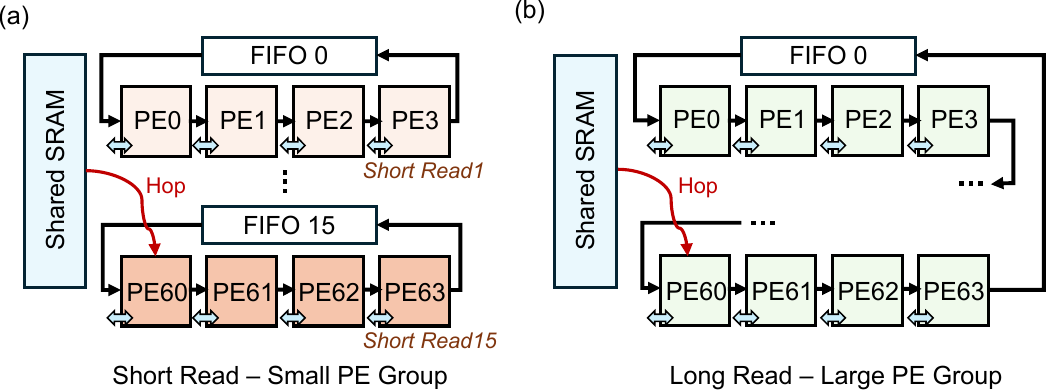}
    \caption{Workload adaptive mapping scheme (a) Short read parallel mapping (b) Long read pipeline mapping}
    \label{fig:s2gmapping}
\end{figure}

The traversal tile resides on the logic base die of the HBM3 stack and couples bit-parallel computation with bank-level parallel data management. As shown in Fig.~\ref{fig:HBM-PNM}(a), we attach one PU to each of the 16 independent HBM channels. Each channel provides 8 bank groups with 32 banks and a 64-bit interface. This static one-to-one mapping maximizes local bandwidth utilization and eliminates contention, with cross-channel communication occurring only for subgraphs spanning multiple channels via a lightweight ring router.

Fig.~\ref{fig:HBM-PNM}(b) details the PU architecture, which contains 64 PEs, a 32 KB input scratchpad, and a 256 KB shared banked SRAM. The input scratchpad stages the linearized reference graph, per-node hop flags, and the pattern bitmasks for the query read, enabling single-cycle streaming into the PE group. To maximize burst locality in HBM, we batch reads by query position.
We store the $i$-th query base and its masks contiguously across the batch, so each PU issues unit-stride HBM bursts. We size the shared banked SRAM at 256 KB to accommodate alignment states for 16 384 nodes per PU. We stripe node states across 32 banks so predecessor gathers spread across banks. Most human variants form short local bubbles, which bounds the active hop working set and increases SRAM hit rate~\cite{10002015global}. To minimize conflicts, we partition this SRAM into 32 banks using hashed bank mapping. Section~\ref{sec:traversal-sensitivity} shows 128\,KB eliminates most spills, and we use 256\,KB to provide headroom for long reads. A 1 KB instruction buffer stores the compact S2G microcode.


The PE micro-architecture illustrated in Fig.~\ref{fig:HBM-PNM}(c) centers on dual 128-bit BPLUs supported by a tiered storage hierarchy comprising a stream register file (SRF), a pattern buffer (PB), and a traceback memory (TBM). Fig.~\ref{fig:HBM-PNM}(d) illustrates the explicit data movement control: micro-instructions \texttt{mv} direct data between local storage and external ports. 

Inside the PE micro-architecture, a four-way input MUX acts as the centralized data entry point to avoid the overhead of general-purpose register files. As illustrated in Fig.~\ref{fig:micro-arch}(a), this selector retrieves operands from the local feedback path (\textit{\textbf{Self}}), the traceback buffer (\textbf{\textcolor[RGB]{25,107,36}{\textit{Replay}}}), the shared banked SRAM (\textbf{\textcolor[RGB]{192,0,0}{\textit{Hop}}}), or adjacent PEs (Neighbor). This design allows the PE to switch between regular systolic streaming and irregular graph-driven memory accesses without architectural stalls. The MUX orchestrates data from the following:

\begin{itemize}
  \item Tier 1 Stream Registers (\textbf{\textit{Self}}). A 384-bit register file holds the working set of current, previous, and dependency vectors. During linear alignment, the MUX selects the \textbf{\textit{Self}} port to create a local feedback loop from the BPLU output, sustaining high throughput for regular updates.
  
\item 
Tier 2 Pattern Buffer. A 256-Byte SRAM provides the pattern masks directly to the BPLU. This capacity allows the hardware to prefetch masks for the next window during current computation to effectively hide load latency.

\item 
Tier 3 Traceback Memory (\textbf{\textcolor[RGB]{25,107,36}{\textit{Replay}}}). A 4 KB circular buffer logs historical direction bits in the background. This capacity covers approximately the same 16 000 steps to accommodate typical segment lengths. When traceback is enabled, the MUX switches to the \textbf{\textcolor[RGB]{25,107,36}{\textit{Replay}}} port to reconstruct paths without global memory access.

\item 
Shared Banked SRAM (\textbf{\textcolor[RGB]{192,0,0}{\textit{Hop}}}). When encountering graph bifurcations, the MUX switches to the \textbf{\textcolor[RGB]{192,0,0}{\textit{Hop}}} port. The state pauses local feedback to fetch non-adjacent predecessor states from the PU-level SRAM and resolves the irregular dependencies inherent to complex topologies.
\end{itemize}

Finally, Fig.~\ref{fig:micro-arch}(b) depicts the reconfigurable link between the two 128-bit BPLU cores. A mode-controlled carry MUX configures the blocks as a 256-bit systolic unit for high-precision or dual 128-bit execution units. This reconfigurability maximizes utilization across diverse sequence lengths.

\subsubsection{Sequence-to-Graph Mapping and Scheduling Scheme}

\label{subsubsection:s2gmapping}

The mapping challenge arises from the structural divergence between linear sequence alignment and graph alignment as shown in Fig.~\ref{fig:genome_sequencing}. Rigid datapaths with fixed buffering cannot fetch non-local predecessors efficiently in pangenomes. Moreover, hardware must adapt to different read lengths: short reads (50 to 300 bp) require high parallelism to manage repeat regions, while long reads (above 10 kb) demand deep computation and high bandwidth to resolve structural variants. Our strategy resolves this by designing a flexible data flow that adapts the hardware configuration to both the read length and the dependency structure.

\textbf{Short-Read Parallel Mapping.}
Short reads on wide pipelines incur high initialization overhead and low utilization. To maximize accelerator throughput, we partition the 64-PE array into 16 independent groups of four PEs as illustrated in Fig.~\ref{fig:s2gmapping}(a). Independent groups process unique read streams in parallel while the PU controller distributes subgraphs and query data to their input scratchpads. This turns the traversal tile into a high-throughput parallel engine that effectively amortizes the control cost of fine-grained tasks. In this parallelized mode, the PEs retain access to the shared banked SRAM to resolve local branches via the \textit{Hop} state.

\textbf{Long-Read Pipeline Mapping.}
Long reads involve deep dependency chains. In this mode, we configure entire 64-PE group as a single deep pipeline, as depicted in Fig.~\ref{fig:s2gmapping}(b). The execution logic remains the same: PEs seamlessly alternate between the \textit{Self} state for linear extensions and the \textit{Hop} state for SRAM-assisted topological traversals. To handle sequences exceeding 10 kb without a linear increase in PE count, the architecture employs a windowed bit-parallel approach with a fixed vector width ($W=128$). This iterative processing provides the necessary computational depth to navigate complex branches and long-range dependencies without stalling, effectively decoupling the physical hardware scale from the total sequence length.

\section{Evaluation}
\subsection{Experimental Setup}

\subsubsection{Baselines}
We evaluate GEN-Graph against three classes of baselines: (i) general-purpose CPU and GPU platforms, (ii) prior S2G accelerators, and (iii) prior large-scale APSP accelerators.
We use an Intel i7-11700K~\cite{intel-i7-specs}, an NVIDIA A100-SXM4~\cite{nvidia-a100-specs}, and an NVIDIA H100-SXM5~\cite{nvidia-h100-specs}.
Table~\ref{tab:hardware-specs} summarizes key parameters. 

For the APSP task, we compare against CPU, GPUs, SOTA PIM method temporal PIM SSSP~\cite{madhavan2021temporal}, and SOTA GPU distributed methods including Partitioned APSP~\cite{djidjev2015all} and Co-Parallel APSP~\cite{sao2021scalable}. Since no SOTA PIM methods directly implement APSP, we estimate the performance of the temporal PIM SSSP~\cite{madhavan2021temporal} to establish a comparable APSP PIM baseline, hereafter referred to as PIM-APSP. For the S2G alignment task, we evaluate the traversal tile against representative implementations across SOTA CPU, GPU, and SOTA accelerator platforms. On CPUs and GPUs, we use SOTA Smith–Waterman based engines, including PASGAL~\cite{jain2019accelerating} and HGA~\cite{feng2021accelerating}, as algorithmic baselines executed on general-purpose hardware. We further compare against two specialized accelerators with distinct design paradigms, namely SeGraM~\cite{cali2022segram}, which follows a PIM approach, and ASGDP~\cite{asgdp}, which adopts a task-specific ASIC design.

\subsubsection{Datasets}
To evaluate GEN-Graph across diverse compute domains, we employ SOTA large-scale datasets for the different compute domains of GEN-Graph: For network analysis workload, we profile the real-world performance using the OGBN-Products graph (2.45M nodes)~\cite{chiang2019cluster}, while architectural sensitivity to topology is assessed via synthetic Newman–Watts–Strogatz (NWS)~\cite{watts1998collective} and Erdős–Rényi (ER)~\cite{erdds1959random} graphs generated by NiemaGraphGen~\cite{moshiri2022niemagraphgen}. For genomic workloads, we further utilize the GRCh38~\cite{grch38} human genome augmented with Genome in a Bottle (GIAB) variations. We then generate a comprehensive read suite using PBSIM2~\cite{ono2021pbsim2} for long reads (PacBio~\cite{pacbio_website}, ONT~\cite{ont_website}) and Mason~\cite{holtgrewe2010mason} for short reads (Illumina~\cite{illumina}), covering a wide range of error profiles. Benchmarks include 100 bp short-reads (LRC-L1: 317.6k reads, MHC1-M1: 497.0k reads) and 10 kbp long-reads (LRC-L2: 3.2k reads, MHC1-M2: 4.9k reads).

\begin{table}[t]
\caption{Hardware Platform Specifications}
\label{tab:hardware-specs}
\renewcommand{\arraystretch}{1.25}
\centering
\scalebox{0.85}{
\begin{tabular}{|l|l|l|l|}
\hline
\textbf{Feature} & \textbf{Intel i7-11700K}~\cite{intel-i7-specs} & \textbf{NVIDIA A100}~\cite{nvidia-a100-specs} & \textbf{NVIDIA H100}~\cite{nvidia-h100-specs} \\
\hline
Architecture & Rocket Lake & Ampere & Hopper \\
Max Freq. & 5.0 GHz (Turbo) & 1.41 GHz (Boost) & 1.98 GHz (Boost) \\
Memory & 64 GB DDR4-3200 & 80 GB HBM2e & 80 GB HBM3 \\
Memory BW & 50 GB/s & 2 039 GB/s & 3.35 TB/s \\
TDP & 125 W & 400 W & 700 W \\
\hline
\end{tabular}
}
\end{table}

\begin{table}[t]
  \centering
  \caption{GEN-Graph Hardware Component Summary}
  \label{tab:hardware-summary}
  \renewcommand{\arraystretch}{1.25}
  \setlength{\tabcolsep}{3pt}
  \scalebox{0.8}{%
    \begin{tabular}{|p{1.8cm}|p{4.6cm}|p{4cm}|}
      \hline
      \multicolumn{1}{|c|}{\textbf{Component}} &
      \multicolumn{1}{c|}{\textbf{Key Feature}} &
      \multicolumn{1}{c|}{\textbf{Function}} \\
      \hline
      \multicolumn{3}{|c|}{\textbf{On-Interposer (2.5D Package)}}\\
      \hline
      HBM3            & 16\,GB capacity; 8-Hi DRAM stack & Traversal computation \\
      \hline
      PCM-FW Die      & 2\,GB capacity; on-tile permutation & PIM-based FW computation \\
      \hline
      PCM-MP Die      & 2\,GB capacity; on-tile min-tree & PIM-based MP merge \\
      \hline
      Logic Base Die  & Dual 64\,KB stream engines & Control, dataflow, format conversion \\
      \hline
      \multicolumn{3}{|c|}{\textbf{On-PCB}}\\
      \hline
      FeNAND          & 16\,TB capacity; ONFI~5.1 & Dense persistent storage \\
      \hline
    \end{tabular}%
  }
\end{table}

\subsubsection{Evaluation Platform}
We developed an in-house cycle-accurate simulator to model the GEN-Graph, validated against RTL synthesis results. With the specialized compiler that maps high-level graph algorithms to tile-specific instructions, the simulator captures the power and performance from the fully end-to-end flow. For the PCM-based matrix tile, we integrate device-level parameters from NeuroSim~\cite{peng2020dnn+} to model analog array operations. For the HBM-based traversal tile, we synthesize the logic die processor using Synopsys Design Compiler with a 28~nm technology library. To enable a fair comparison with the 7~nm GPU baselines, we scale all area and power results to the 7~nm node using established scaling models~\cite{stillmaker2017scaling}.

\begin{table}[t]
    \centering
    \caption{Hardware configurations of SLC PCM~\cite{wu2024novel}}
    \label{tab:pcm}
    \begingroup
    \renewcommand{\arraystretch}{1.25}
    \scalebox{0.89}{
        \begin{tabular}{|l|c|l|c|}
            \hline
            \multicolumn{4}{|c|}{\textbf{SLC PCM Die Parameters}} \\ 
            \hline
            Read Energy/bit & 0.05 pJ & Write Energy/bit & 0.56 pJ \\
            \hline
            Read/Write Latency & 2 ns/20 ns & Die Capacity & 2 GB \\
            \hline
            Clock Cycle & 2 ns (500 MHz) & Reset Pulse & 40 \textmu A @ 0.70 V \\
            \hline
            Resistance (LRS) & $\approx 30$ k$\Omega$ & On/Off Ratio & 150$\times$ \\
            \hline
            \multicolumn{4}{|c|}{\textbf{PCM Organization}} \\

            \hline
            Organization & \multicolumn{3}{l|}{\begin{tabular}[l]{@{}l@{}}1024 by 1024 cells per unit;\\ 130 units per tile; 128 tiles per die.\end{tabular}} \\
            \hline
        \end{tabular}
    }
    \endgroup
\end{table}

\begin{table}[t]
    \centering
    \caption{Hardware configurations of HBM3 DRAM}
    \label{tab:dram}
    \renewcommand{\arraystretch}{1.25}
    \scalebox{0.72}{
        \begin{tabular}{|l|c|l|c|}
            \hline
            \multicolumn{4}{|c|}{\textbf{HBM3 DRAM Device Parameters}} \\ 
            \hline
            \#layers & 8-Hi & Technology Node & 10nm (1Ynm or 1Znm) \\
            \hline
            Bank Capacity & 256 Mb & Bank Area & 0.219 mm\textsuperscript{2} \\
            \hline
            Channel Row Buffer & 16 Kb & Read/Write Energy/bit & 0.4 pJ / 0.45 pJ \\
            \hline
            Chip Area & 121 mm\textsuperscript{2} & Read/Write Latency/ns & 10-20 ns \\
            \hline
            \multicolumn{4}{|c|}{\textbf{HBM3 System Parameters}} \\

            \hline
            Organization & \multicolumn{3}{l|}{\begin{tabular}[l]{@{}l@{}}16 channels per chip (64-bit I/O per channel);\\ 32 banks per channel.\end{tabular}} \\

            \hline
            \multicolumn{4}{|c|}{\textbf{Processing Element (PE)}} \\
            \hline
            Compute Unit Array & Dual-BPLU (2$\times$128b) & Stream Register File & 384b\\
            \hline
            Pattern Buffer & 256 B & Traceback Memory & 4 KB\\
            \hline
            \multicolumn{4}{|c|}{\textbf{Processing Unit (PU)}} \\
            \hline
            \#PEs & 64 & Shared Banked SRAM & 256 KB \\
            \hline
           Instruction Buffer & 1KB  & Ring Router & 128 GB/s/link \\
            \hline
            \multicolumn{4}{|c|}{\textbf{HBM3 DRAM NMP Processor}} \\
            \hline
            Basic & \multicolumn{3}{l|}{28 nm process; 0.7 V supply; 113 mm\textsuperscript{2} die area; INT8/INT32 format} \\
            \hline
            \#PUs & 16 & SRAM Capacity & 4 MB (Distributed) \\
            \hline
            Peak Bandwidth & 819.2 GB/S & Peak Power & 8.52 W \\
            \hline
        \end{tabular}
    }
\end{table}

\subsubsection{GEN-Graph configuration}
\label{config}

Table~\ref{tab:hardware-summary} lists the GEN-Graph configuration, which integrates HBM3-backed traversal tile and PCM-based matrix tile on a 2.5D interposer, with FeNAND providing off-package capacity for large graphs and persistent data. Table~\ref{tab:pcm} summarizes the PCM in-memory processor used by the matrix tile. Each tile operates on $1024 \times 1024$ arrays, enabling fully in-place execution. Table~\ref{tab:dram} summarizes the HBM3 near-memory processor used by the traversal tile: 16 HBM channels, each with 32 banks, provide high bandwidth for irregular accesses. We assess the impact of these architectural parameters in the scalability tests for both the GEN-Graph matrix tile (Section~\ref{sec:apsp_result}) and traversal tile (Section~\ref{sec:s2g_result}).


\subsection{GEN-Graph Performance on APSP Workloads}
\label{sec:apsp_result}

\begin{figure}[t]
    \centering
        \includegraphics[width=1\linewidth]{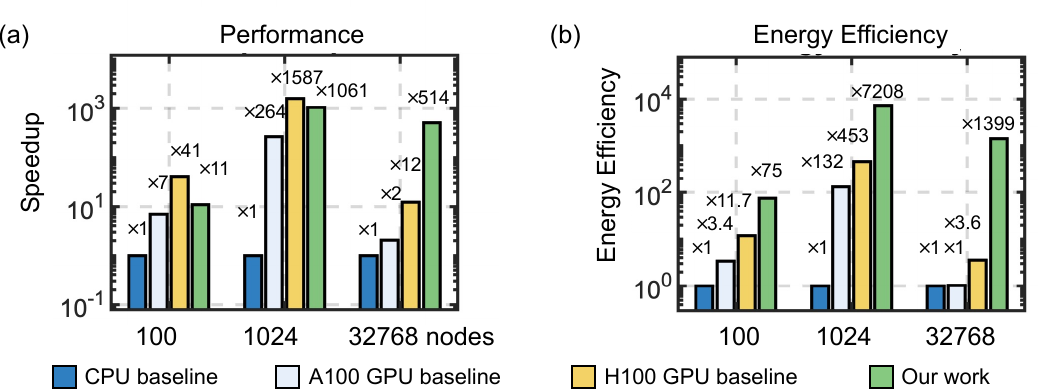}
    \caption{GEN-Graph matrix tile vs. CPU, A100 GPU and H100 GPU baselines across graph sizes (a) Speedup (b) Energy efficiency}
    \label{fig:baseline}
\end{figure}

\begin{figure}[t]
    \centering
    \includegraphics[width=1\linewidth]{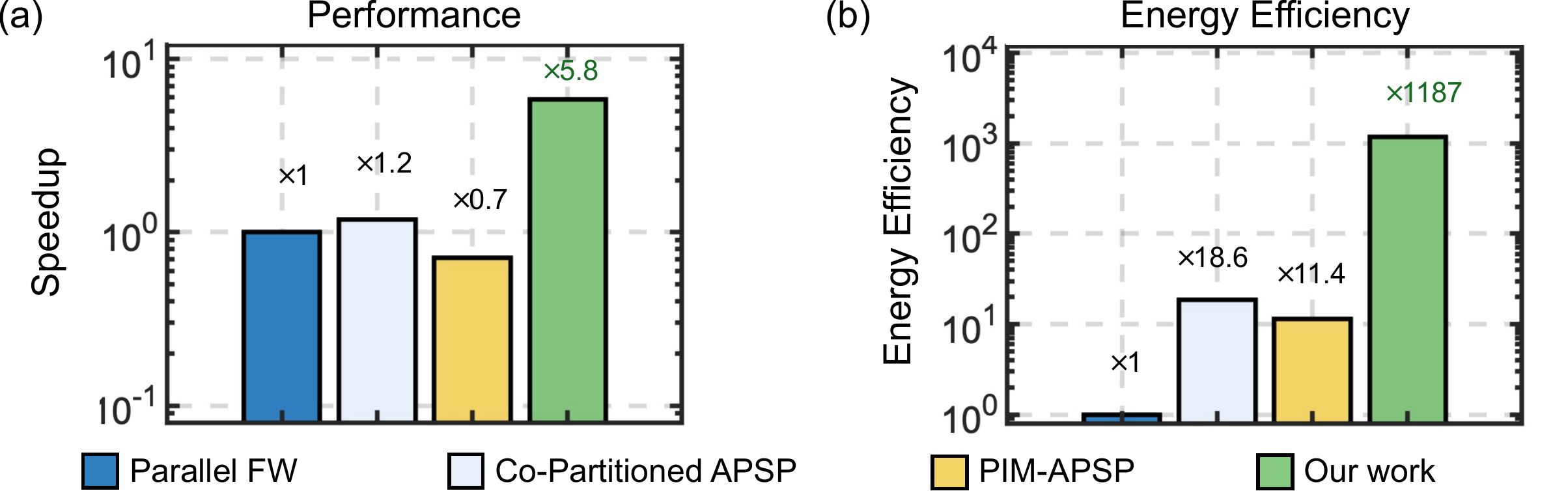}
    \caption{GEN-Graph matrix tile vs. PIM-APSP~\cite{madhavan2021temporal}, Partitioned APSP~\cite{djidjev2015all}, and Co-Parallel APSP~\cite{sao2021scalable} running APSP on OGBN-Products dataset~\cite{chiang2019cluster} (a) Speedup (b) Energy efficiency}
    \label{fig:existingwork}
\end{figure}

\begin{figure}[t]
    \centering
    \includegraphics[width=1\linewidth]{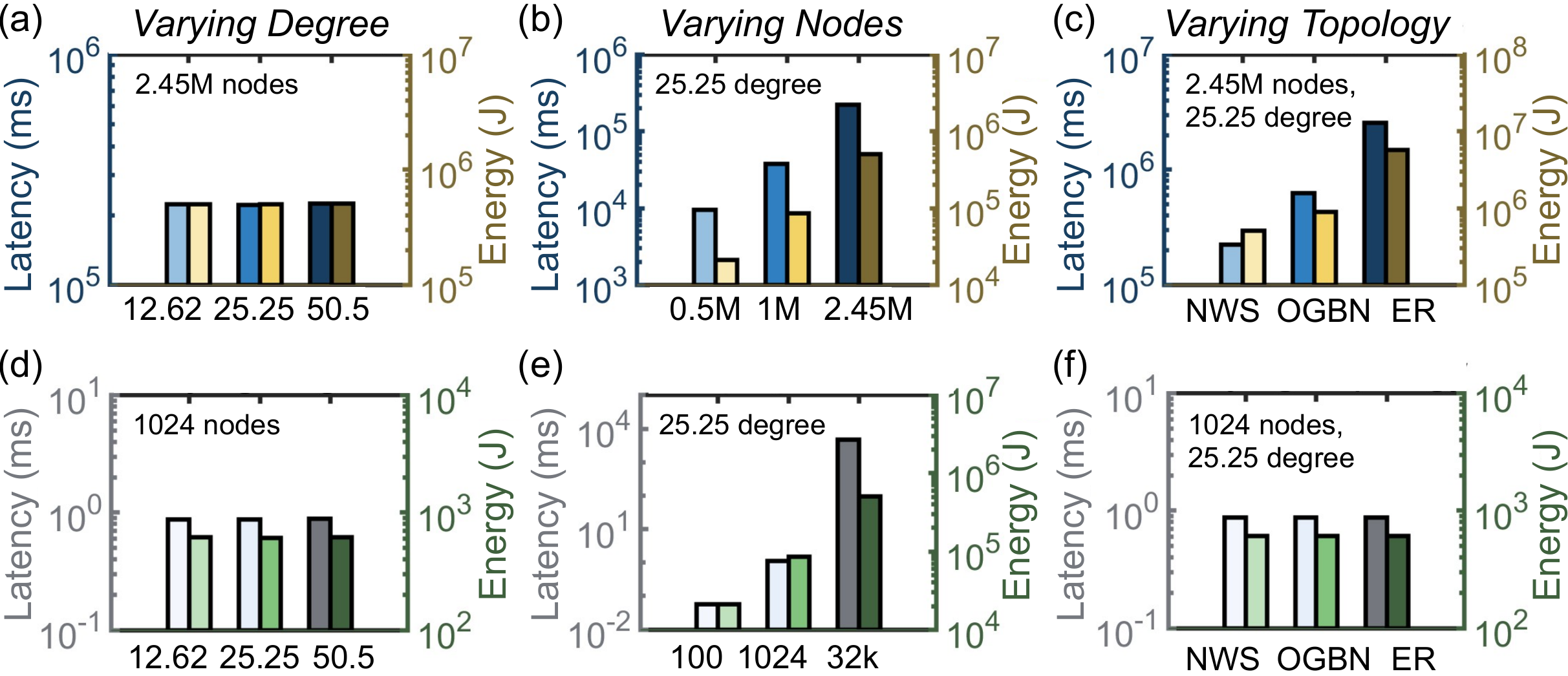}
    \caption{Scalability of matrix tile across (a) Degree (b) Size and (c) Graph topologies; and H100 across (d) Degree (e) Size and (f) Graph topologies. Topologies include clustered (NWS), real (OGBN), and random (ER)}
    \label{fig:apsp-scale}
\end{figure}

Fig.~\ref{fig:baseline} shows a comparison of GEN-Graph matrix tile running APSP workloads to CPU and GPU (A100 and H100) baselines using graphs with 100, 1\,024, and 32\,768 nodes synthesized using NiemaGraphGen~\cite{moshiri2022niemagraphgen}. These sizes avoid memory bottlenecks on single-node hardware. At 1\,024 nodes, matrix tile delivers $1\,061\times$ speedup and $7\,208\times$ energy efficiency over CPU. At 32\,768 nodes, it outperforms H100 by $42.8\times$ in speed and $392\times$ in energy. Matrix tile's performance gains grow with graph size due to enhanced parallelism and in-memory execution. This performance gap widens dramatically with larger graph sizes because conventional systems are overwhelmed by $O(n^3)$ data movement. This massive data transfer saturates memory bandwidth that matrix tile inherently avoids. Fig.~\ref{fig:existingwork} then compares GEN-Graph matrix tile, running APSP workloads, with SOTA accelerator PIM-APSP~\cite{madhavan2021temporal}, as well as with GPU cluster methods -- Partitioned APSP~\cite{djidjev2015all} and Co-Parallel APSP~\cite{sao2021scalable}. On OGBN-Products (2.45M nodes) dataset~\cite{chiang2019cluster}, we estimate their performance from reported scaling trends. Matrix tile outperforms both, achieving $5.8\times$ speedup over Co-Parallel APSP~\cite{sao2021scalable} and $1\,186\times$ energy savings over Partitioned APSP~\cite{djidjev2015all}. While PIM-APSP improves energy efficiency by $11.4\times$, it slows down performance to $70\%$ of the baseline. The advantage of matrix tile comes from removing inter-GPU communication at large-scale, multi-node APSP solutions.

\textbf{Scalability Analysis of Matrix Tile.}
We compare the scalability of GEN-Graph matrix tile and the H100 GPU baseline by varying graph degrees, size, and topology. In Fig.~\ref{fig:apsp-scale}(a) and (d), both systems maintain stable performance as a function of degree, suggesting edge count has a limited effect on exact APSP when there is enough memory. In Fig.~\ref{fig:apsp-scale}(b) and (e), GEN-Graph matrix tile scales linearly to 2.45M nodes, while H100 exhibits rising latency and superlinear energy growth beyond $10^3$ nodes due to communication overhead. H100 is limited by the $O(n^2)$ distance matrix overwhelms its caches, whereas matrix tile's in-situ computation preserves locality. In Fig.~\ref{fig:apsp-scale}(c) and (f), GEN-Graph matrix tile achieves better efficiency on clustered and real-world graphs than on random ones, benefiting from structural locality and fewer partitioning boundaries, while H100 remains largely topology-insensitive. This topology-awareness is a direct result of our partitioning algorithm, as clustered graphs like NWS produce smaller boundary sets, reducing the workload of the computationally dominant boundary-graph APSP step. GEN-Graph matrix tile provides competitive scalability without the immense hardware cost and overhead of GPU clusters. Partitioned-FW~\cite{djidjev2015all} uses 2 560 GPUs for a 1.9M-node graph but hits synchronization and memory walls. Co-ParallelFW~\cite{sao2021scalable} achieves only 45\% weak-scaling efficiency on a 300K-node graph. Overall, GEN-Graph matrix tile avoids the complexity, synchronization, and energy overheads of distributed GPU clusters.

We further compared to prior GPU-based APSP systems, GEN-Graph matrix tile offers competitive scalability with far lower hardware cost. Partitioned-FW~\cite{djidjev2015all} scales to 2\,560 GPUs and solves 1.9M-node graphs in six minutes, but faces synchronization and memory bottlenecks at scale. Co-ParallelFW~\cite{sao2021scalable} achieves $4.6\times$ speedup from 16 to 256 GPUs per node with $45\%$ efficiency on 300K nodes via weak scaling. In summary, GEN-Graph matrix tile leverages topology awareness to scale efficiently, avoiding the synchronization and energy penalties that restrict distributed GPU clusters.

\textbf{Sensitivity Analysis of Matrix Tile.} We examine how the effective GEN-Graph matrix tile capacity \(N\) affects end-to-end APSP performance on OGBN-Products. 
\begin{wrapfigure}{r}{0.2\textwidth}
  \begin{center}
    \vspace{-10pt} 
    \includegraphics[width=0.19\textwidth]{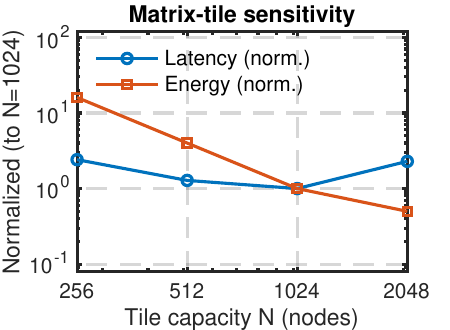}
    \vspace{-15pt} 
  \end{center}
  \caption{Sensitivity to matrix tile capacity $N$}
  \label{fig:matrix-sensitivity}
\end{wrapfigure}
In Fig.~\ref{fig:matrix-sensitivity}, we vary \(N \in \{256,512,1024,2048\}\) and normalize latency and energy to the \(N=1024\) design point. The sensitivity curve reflects three competing effects: smaller tiles increase recursive partitioning and cross-component merge overhead, while larger tiles reduce on-die parallelism (scaling approximately as \(1/N^{2}\)) and incur RC-limited frequency degradation modeled as \((N/1024)^{\alpha}\) with \(\alpha=1.3\). As a result, latency follows a convex trend and is minimized at \(N=1024\) with normalized latency \(1.0\times\). In comparison, \(N=256\), \(512\), and \(2048\) incur respectively \(2.41\times\), \(1.28\times\), and \(2.29\times\) higher latency. We therefore adopt \(1024\times1024\) PCM arrays as the default matrix tile configuration for the remainder of this work.

\subsection{GEN-Graph Performance on S2G Workload}
\label{sec:s2g_result}
\begin{figure}[t]
    \centering
    \includegraphics[width=1\linewidth]{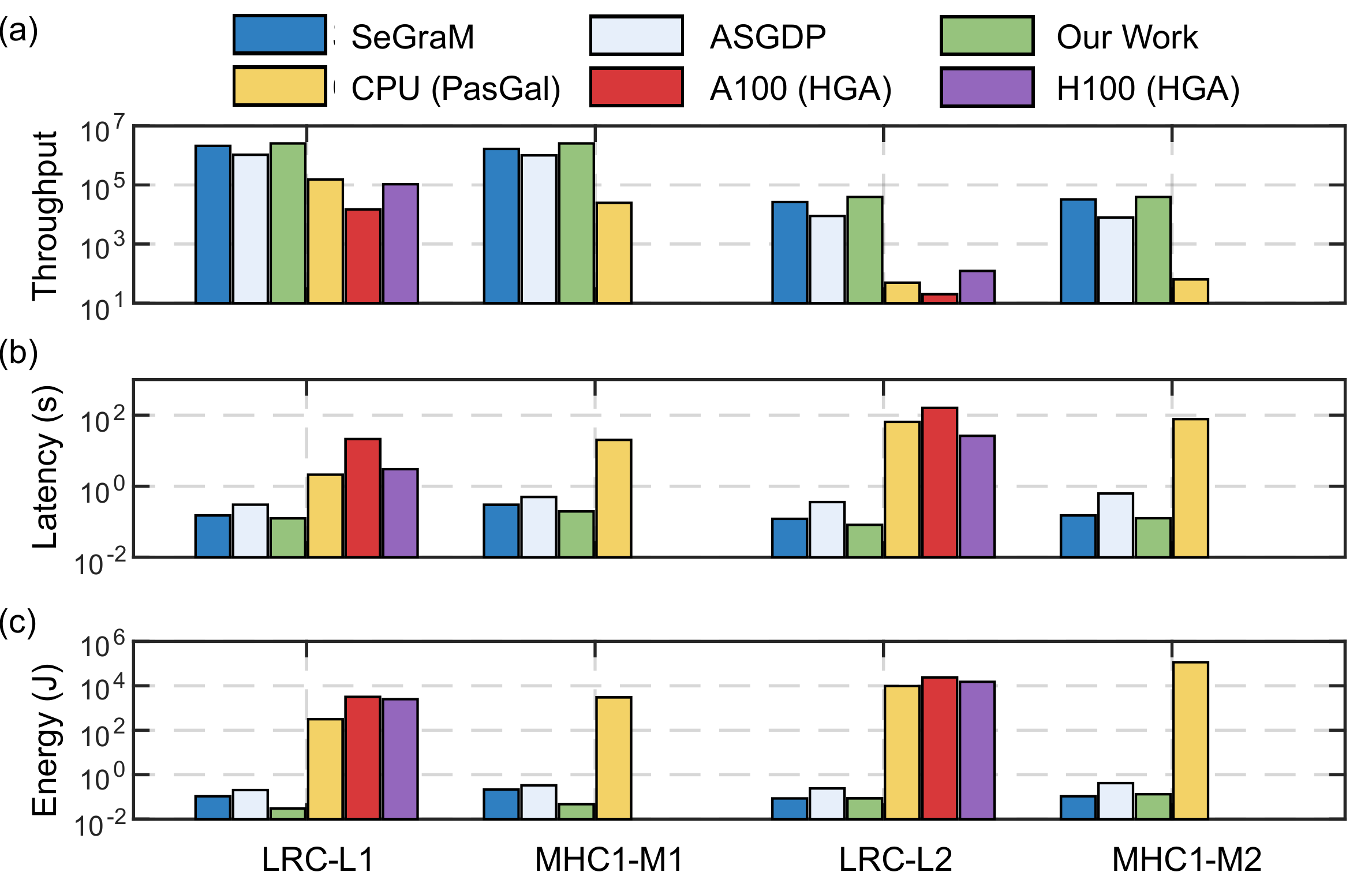}
    \caption{GEN-Graph traversal tile vs. SOTA CPU, GPU and HW accelerators on S2G including SeGraM~\cite{cali2022segram}, ASGDP~\cite{asgdp}, PasGal~\cite{jain2019accelerating} and HGA~\cite{feng2021accelerating} (a) Throughput (b) Latency  (c) Energy}
    \label{fig:S2G-performance}
\end{figure}

We evaluate the end-to-end throughput of GEN-Graph traversal tile using four representative workloads SeGraM~\cite{cali2022segram}, ASGDP~\cite{asgdp}, PasGal~\cite{jain2019accelerating} and HGA~\cite{feng2021accelerating}, targeting the short and long read workloads. In Fig.~\ref{fig:S2G-performance}(a), the GEN-Graph traversal tile achieves the best throughput across all benchmarks. On the two short read workloads (LRC-L1 and MHC1-M1), it reaches about 2.56 million reads/s and improves throughput over SeGraM~\cite{cali2022segram} by 21\% and 55\%. It also outperforms ASGDP~\cite{asgdp} by 2.44$\times$ and 2.56$\times$. Here, the gain comes from sustaining higher PU utilization under irregular dependencies by the tiered storage hierarchy. This co-designed approach enables GEN-Graph to sustain high hardware utilization by overlapping in-place computation with hierarchical data streaming.

\begin{figure}[t]
    \centering
    \includegraphics[width=1\linewidth]{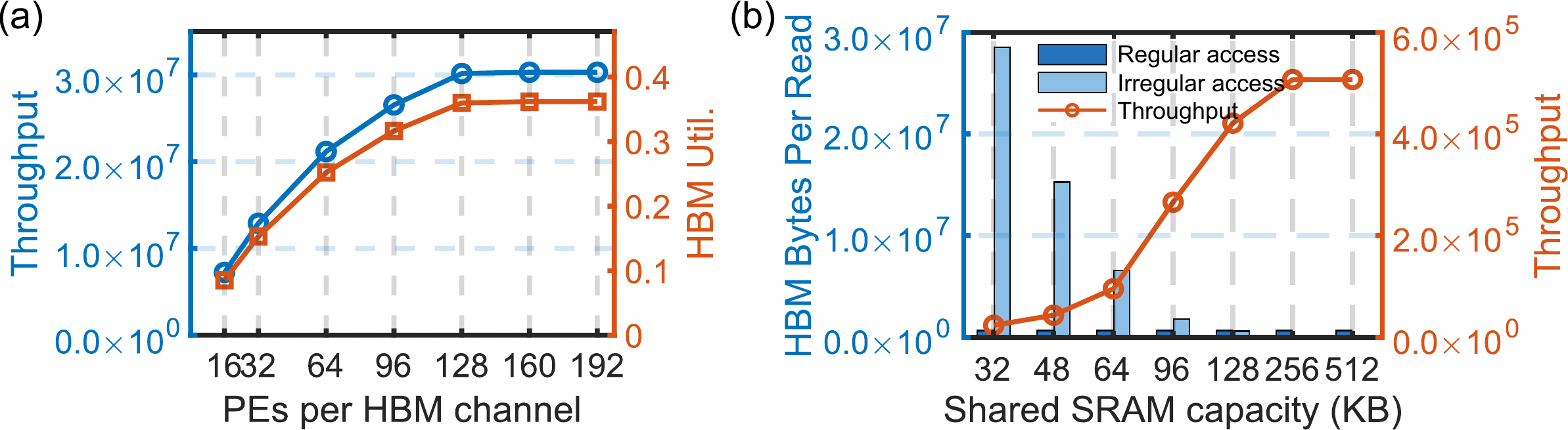}
    \caption{Sensitivity to (a) PE scalability and (b) Shared SRAM capacity}
    \label{fig:ttsensitivity}
\end{figure}

In Fig.~\ref{fig:S2G-performance}(b), the GEN-Graph traversal tile is 171$\times$ faster on LRC-L1 and 1963$\times$ faster on LRC-L2, while reducing energy by 1.05$\times10^{5}$ and 2.79$\times10^{5}$ compared to same workloads running on A100 GPU~\cite{nvidia-a100-specs}. The GPU runtime grows sharply on long reads because DP tables expand and exceed on-chip cache capacity.
This forces frequent global memory accesses and exposes DRAM latency.
GEN-Graph avoids this mode by storing the live dependency state in SRAM and executing the inner DP loop on the logic die. Fig.~\ref{fig:S2G-performance}(c) shows that GEN-Graph also improves energy efficiency. It reduces energy versus SeGraM~\cite{cali2022segram} by 3.53$\times$ (LRC-L1) and 4.51$\times$ (MHC1-M1), and versus ASGDP~\cite{asgdp} by 6.72$\times$ and 7.06$\times$. This comes from data movement. GPU executions repeatedly move DP tables through high-power memory and interconnects.
GEN-Graph keeps the DP working set on-package and streams inputs through a narrow, predictable path.

\textbf{Scalability Analysis of Traversal Tile.}
Fig.~\ref{fig:S2G-performance}(b) exposes the performance comparison across different read length workloads include 100 bp short-reads (LRC-L1: 317.6k reads, MHC1-M1: 497.0k reads) and 10 kbp long-reads (LRC-L2: 3.2k reads, MHC1-M2: 4.9k reads). For 100\,bp reads, the compiler partitions PUs into independent parallel groups and sustains 2.56 million reads/s on LRC-L1. When scaling to long reads in LRC-L2, it switches to a deep pipeline and sustains 39.3 thousand reads/s on LRC-L2. In contrast, the GPU baselines show exponential latency growth on long reads, suffering from limited on-chip cache capacity. This shows GEN-Graph's configurable mapping strategy maintains high utilization across varied read lengths.

\textbf{Sensitivity Analysis of Traversal Tile.}
\label{sec:traversal-sensitivity}
We evaluate the sensitivity of the traversal tile to PE density and local memory capacity to validate our allocation strategy on the logic die.

\textit{PE Density Effects.} 
Fig.~\ref{fig:ttsensitivity}(a) sweeps the number of PEs per HBM channel from 16 to 192 with a mixed workload of short and long reads. The left axis in blue shows aggregate throughput, while the right axis in red shows HBM bandwidth utilization. Aggregate throughput increases from 7.2M reads/s at 16 PEs to 21.2M reads/s at 64 PEs, corresponding to a \(2.9\times\) improvement. Increasing the density to 96 PEs further raises throughput to 26.5M reads/s. Between 128 and 192 PEs, throughput saturates near 30.2M reads/s, improving by less than 1.1\% despite a \(1.5\times\) increase in PE count. This saturation coincides with HBM bandwidth utilization increasing from 0.09 at 16 PEs to approximately 0.36 at 128 PEs, and remaining nearly constant thereafter. The selected PE density 64 therefore lies near the knee of the throughput saturation curve, capturing most attainable performance without over-provisioning compute resources.

\textit{Shared SRAM Capacity Effects.}
Fig.~\ref{fig:ttsensitivity}(b) sweeps the shared SRAM from 32\,KB to 512\,KB using long reads. The bars on the left axis in blue report HBM bytes per read split into topology streaming (regular) and traversal accesses (irregular), while the right axis in red plots kernel-only throughput versus shared SRAM capacity. We use long reads to expose memory-bound scaling, since short-read dependencies usually fit within minimal local storage. At 32 KB, limited capacity causes dependency state spills that generate 29.2 MB of HBM traffic and restrict throughput to 23.9K reads/s. Expanding the capacity to 128 KB reduces total traffic to 1.3 MB by dropping state spills from 28.5 MB to 0.62 MB. This results in a $17.6\times$ throughput gain to 421K reads/s. Topology streaming stays constant at 0.68\,MB. Once state spills remain trivial, throughput saturates at 507K reads/s for both 256\,KB and 512\,KB. This trend indicates that 128 KB shared banked SRAM is sufficient to store the active working set; beyond that, performance is bounded by topology streaming, not by capacity misses.



\subsection{Area and Power Analysis}

GEN-Graph achieves these gains while preserving exact DP updates, with no pruning or approximation.
Across our workloads, it exceeds A100~\cite{nvidia-a100-specs} by $42.8\times$ in APSP speed and exceeds H100~\cite{nvidia-a100-specs} by up to $1963\times$ in S2G latency, while reducing energy by up to $2.79\times10^{5}$.

\textbf{Area and Power Breakdown.}
Table~\ref{tab:combined-breakdown} details the physical parameters of the compute tiles based on synthesis and simulation results. In the matrix tile, peripherals, including row drivers, sense amplifiers, and global wires, occupy over 81\% of area to meet PCM operation constraints.
During updates, PCM subarrays dominate power and account for 80.6\% of consumption. In the traversal tile, the PE array and shared SRAM account for 94\% of area.
Power splits between compute (18.4\,mW) and local memory access (19.4\,mW), consistent with a state-heavy inner loop. The full system integrates these tiles with HBM3 (16\,GB) adds $8.6$\,W and $121\,mm^2$~\cite{park2024attacc}; FeNAND (16\,TB) adds $6.4$\,W across $3000\,mm^2$; the SM2508 controller adds $3.5$\,W within a $225$\,mm$^2$ BGA package. The total power of $\sim18.5$\,W remains significantly lower than high-end GPUs such as the NVIDIA H100~\cite{nvidia-h100-specs}, which consume up to $700$\,W under peak workloads~\cite{luo2024benchmarking} with higher latency.

\begin{table}[t]
\caption{Area and Power Breakdown by Tile Component}
\label{tab:combined-breakdown}
\centering
\renewcommand{\arraystretch}{1.25}
\setlength{\tabcolsep}{3pt}
\resizebox{\columnwidth}{!}{%
\begin{tabular}{|l|cc|cc|}
\hline
\multicolumn{5}{|c|}{\textbf{Matrix Tile (PCM Unit Level)}} \\
\hline
\textbf{Component} & \multicolumn{2}{c|}{\textbf{PCM-FW Unit}} & \multicolumn{2}{c|}{\textbf{PCM-MP Unit}} \\
\cline{2-5}
& \textbf{Area ($\mu$m$^2$)} & \textbf{Power (mW)} & \textbf{Area ($\mu$m$^2$)} & \textbf{Power (mW)} \\
\hline
PCM Subarray & 3 288 (13.8\%) & 557 (80.6\%) & 3 288 (13.6\%) & 557 (80.6\%) \\
Permutation Unit & 917.3 (3.9\%) & 0.59 ($<$0.1\%) & --- & --- \\
Min Comparator & --- & --- & 1 268 (5.3\%) & 0.68 (0.1\%) \\
Controller & 5.9 ($<$0.1\%) & $<$0.01 ($<$0.1\%) & 5.9 ($<$0.1\%) & $<$0.01 ($<$0.1\%) \\
Peripherals* & 19 610 (82.3\%) & 133.3 (19.3\%) & 19 610 (81.1\%) & 133.3 (19.3\%) \\
\hline
\textbf{PCM Total} & \textbf{23 821} & \textbf{690.9} & \textbf{24 172} & \textbf{691.0} \\
\hline
\hline
\multicolumn{5}{|c|}{\textbf{Traversal Tile (HBM Logic Die - Per PU)}} \\
\hline
\textbf{Component} & \multicolumn{2}{c|}{\textbf{Area ($\mu$m$^2$)}} & \multicolumn{2}{c|}{\textbf{Power (mW)}} \\
\hline
PE Array (64 PEs) & \multicolumn{2}{c|}{24 400 (49.08\%)} & \multicolumn{2}{c|}{18.4} \\
Shared SRAM      & \multicolumn{2}{c|}{22 400 (44.99\%)} & \multicolumn{2}{c|}{19.4} \\
Instr. Buffer     & \multicolumn{2}{c|}{91 (0.18\%)}      & \multicolumn{2}{c|}{1.36} \\
Input Scratchpad  & \multicolumn{2}{c|}{2 860 (5.75\%)}   & \multicolumn{2}{c|}{2.91} \\
\hline
\textbf{PU Total} & \multicolumn{2}{c|}{\textbf{50 000}}  & \multicolumn{2}{c|}{\textbf{41}} \\
\hline

\end{tabular}%
}
\vspace{2pt}

\footnotesize{*Peripherals include row drivers, sense amplifiers, and global wires.}
\end{table}

\textbf{Overhead Analysis.}
We evaluate the hardware cost of specialized computation by isolating logic area and power against the baseline memory arrays. Integrating the permutation unit and min-comparator adds only 3.9\% and 5.3\% to the unit area, respectively. The power overhead for this logic remains below 0.1\% compared to the energy required for PCM cell operations. On the traversal tile, the 16 PUs utilize available silicon space on the HBM logic die without increasing the package footprint. By isolating state feedback, mask prefetching, and traceback into specialized buffers, the architecture reduces the area and power overhead associated with complex address generation. Overall, GEN-Graph provides accelerator-level throughput with array-dominated area and power.

\section{Conclusion}

GEN-Graph resolves the fundamental compute-memory divergence in graph-based DP through a heterogeneous 2.5D PIM architecture. By matching hardware specialization to algorithmic structure, GEN-Graph integrates high-density PCM-based matrix tiles for compute-bound patterns and low-latency HBM-based traversal tiles for memory-bound genomic traversals. Our hardware-software co-design utilizes recursive partitioning and reconfigurable bit-parallel logic to ensure exact computation while maximizing resource utilization across billion-node graphs. Our compiler-driven mapping further maximizes hardware occupancy by overlapping bit-serial PCM updates with hierarchical data streaming. Evaluation results demonstrate that GEN-Graph achieves a $42.8\times$ speedup and $392\times$ energy efficiency improvement over the NVIDIA H100 GPU for APSP, while exceeding prior PIM accelerators Temporal PIM SSSP~\cite{madhavan2021temporal} by $8.3\times$ and $104\times$. For S2G alignment, the architecture sustains a throughput of up to 2.56 million reads/s, outperforming SeGraM by up to 55\% and SOTA ASIC ASGDP~\cite{asgdp} by up to $2.56\times$. GEN-Graph establishes a scalable foundation for exact DP acceleration in large-scale genomic and network applications.

\section*{Acknowledgment}
This work was supported by PRISM and CoCoSys—centers in JUMP 2.0, an SRC program sponsored by DARPA (SRC grant number - 2023-JU-3135); the U.S. DOE DeCoDe project \#84245 at PNNL; and NSF grants \#2003279, \#1911095, \#2112167, \#2052809, \#2112665, \#2120019, \#2211386.



\bibliographystyle{IEEEtran}
\bibliography{Ref}
\raggedbottom

\vspace{-1.4 cm}

\begin{IEEEbiography}[{\includegraphics[width=0.9in,height=1.2in,clip,keepaspectratio]{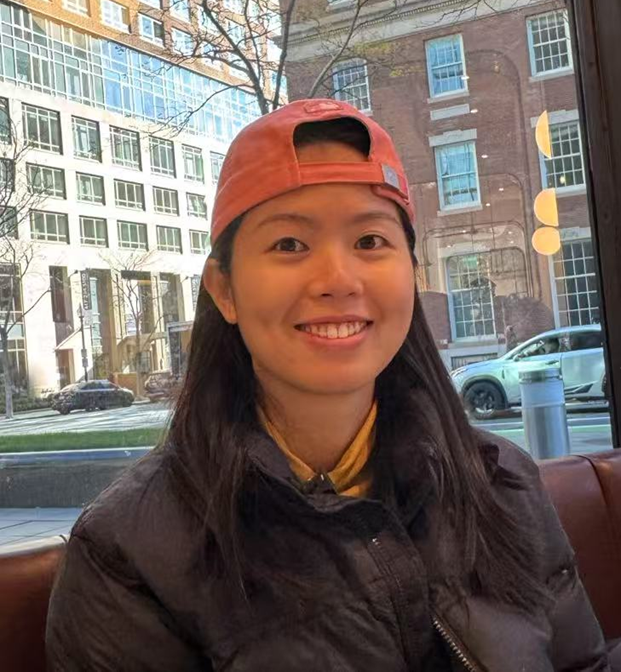}}]{Yanru Chen}
is currently a Ph.D. student in Electrical and Computer Engineering at the University of California San Diego, La Jolla, CA, USA. She received the M.S. degree in Electronic Information (Intelligent Manufacturing) from Tsinghua University, Beijing, China, in 2024, and the B.E. degree in Electronic Science and Technology from Jilin University, Changchun, China, in 2021. Her research interests include SW/HW co-design, processing-in-memory, in-storage processing, graph analytics, hyperdimensional computing, and database search. 

\end{IEEEbiography}

\vspace{-1.6cm}

\begin{IEEEbiography}[{\includegraphics[width=0.9in,height=1.2in,clip,keepaspectratio]{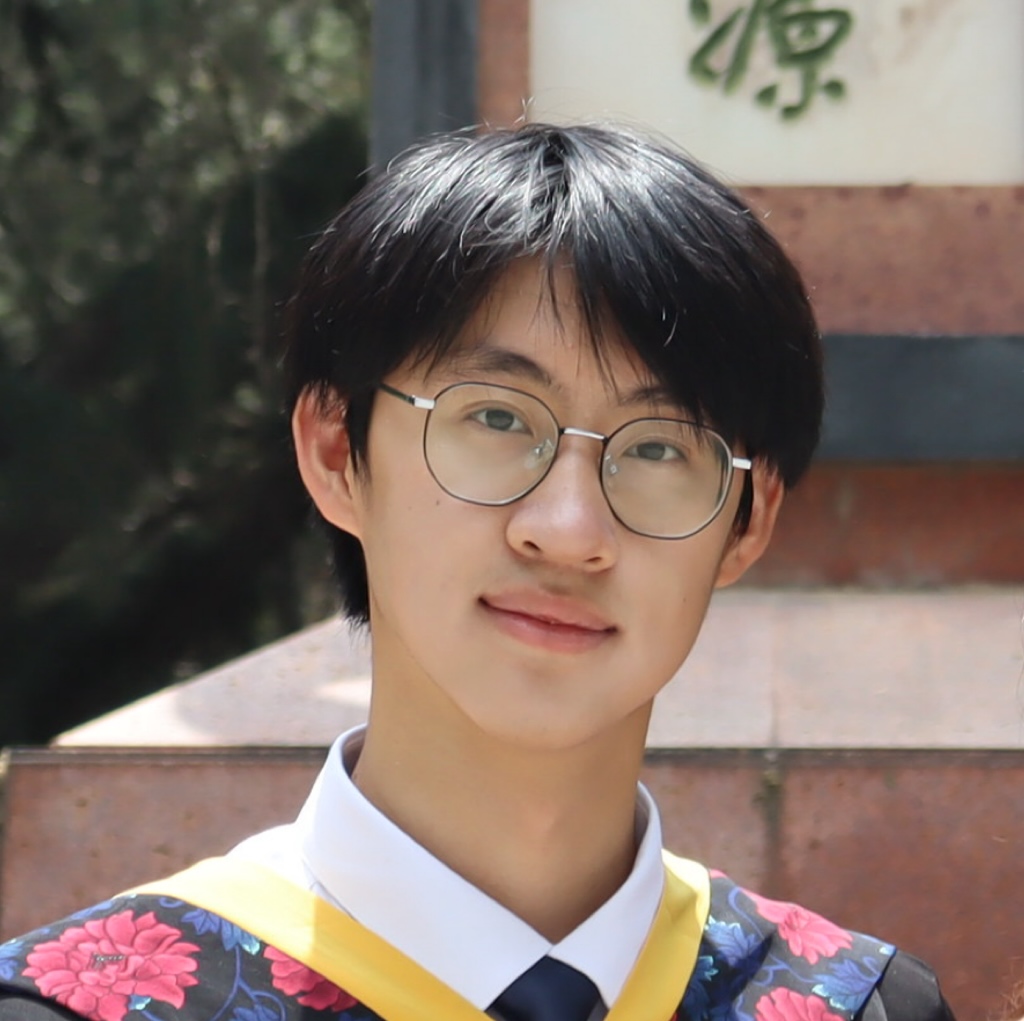}}]{Runyang Tian}
is currently an M.S. student in Electrical and Computer Engineering at the University of California San Diego, La Jolla, CA, USA. He received the B.E. degree in Microelectronic Science and Engineering from Xi'an Jiaotong University, Xi'an, China, in 2024. His research interests include memory-centric computation and domain-specific accelerators.
\end{IEEEbiography}

\vspace{-2cm}

\begin{IEEEbiography}[{\includegraphics[width=0.9in,height=1.2in,clip,keepaspectratio]{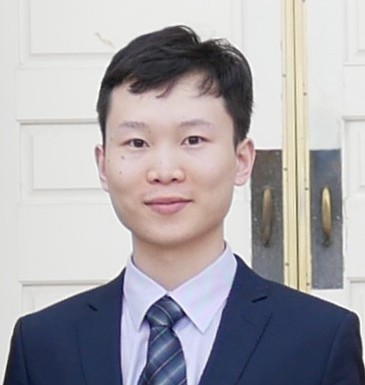}}]{Zheyu Li} received a B.Sc degree in Electrical Engineering and a Ph.D. in Computer Science and Engineering from Pennsylvania State University, University Park, PA, USA, in 2017 and 2023 respectively. He currently is a postdoc in the Energy Efficiency Lab at the University of California, San Diego (UCSD). His research interests lie in hardware and software acceleration of data-intensive emerging workloads, including bioinformatics applications, brain-inspired computing, near-memory computing, and FPGA-based architectures.
\end{IEEEbiography}

\vspace{-1.4cm}

\begin{IEEEbiography}[{\includegraphics[width=0.9in,height=1.2in,clip,keepaspectratio]{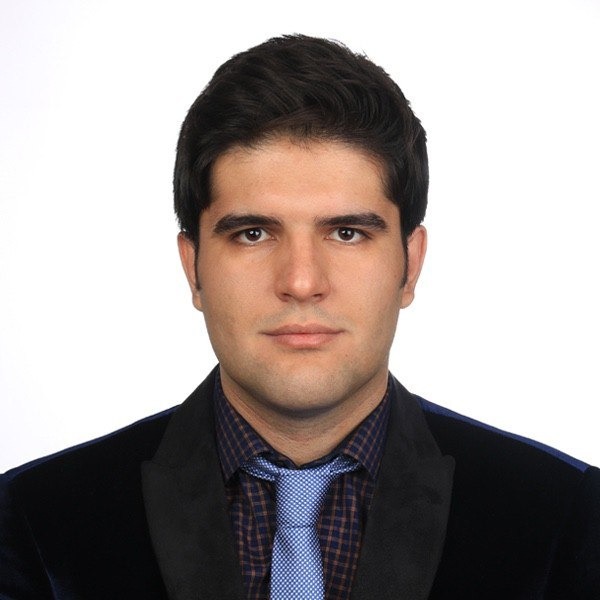}}]{Mahbod Afarin}
is currently a postdoctoral researcher in the Department of Computer Science at the University of California, San Diego, La Jolla, CA, USA. He received his Ph.D. in Computer Science from the University of California, Riverside, CA, USA, in 2024, and his M.Sc. in Computer Engineering from Sharif University of Technology, Tehran, Iran, in 2018. His research interests include hardware accelerators, compiler optimization, and graph analytics.
\end{IEEEbiography}

\vspace{-1.6cm}

\begin{IEEEbiography}[{\includegraphics[width=0.9in,height=1.2in,clip,keepaspectratio]{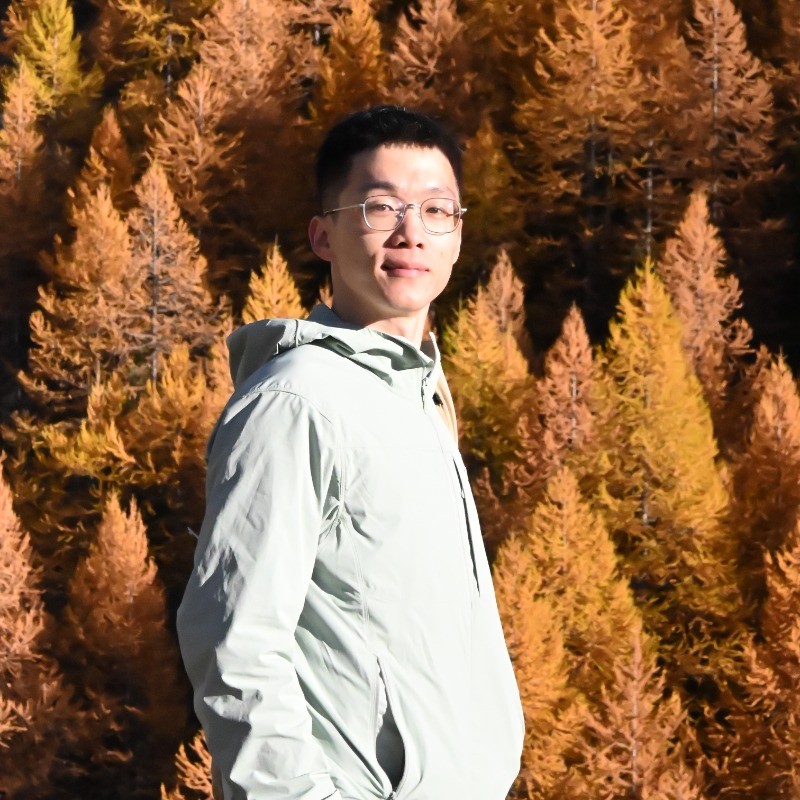}}]{Weihong Xu}
is currently a Postdoc in École polytechnique fédérale de Lausanne. He received PhD degree in Computer Engineering at the University of California San Diego, La Jolla, CA, USA. Before joining UCSD, he received B.E. and M.E. degrees from Southeast University. His research focuses on next-generation computing architectures for efficient and reliable AI systems, including: a) RISC-V-based AI chip, b) computer architecture and EDA co-design, c) near-data computing, d) LLM and AI acceleration.

\end{IEEEbiography}

\vspace{-1.6cm}

\begin{IEEEbiography}[{\includegraphics[width=0.9in,height=1.2in,clip,keepaspectratio]{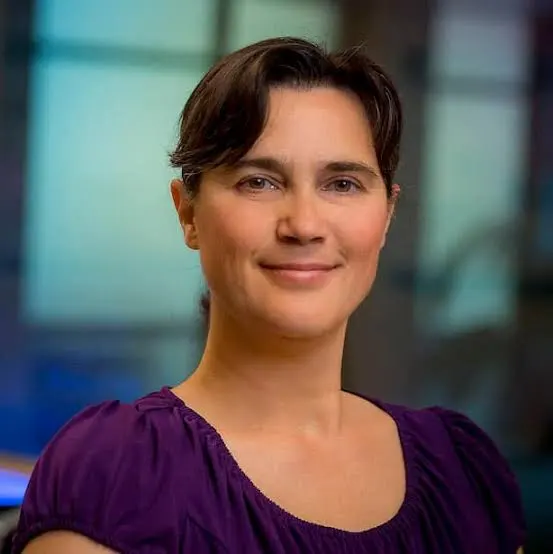}}]{Tajana Rosing}
received her Ph.D. degree from Stanford University, Stanford, CA, USA, in 2001. She is a Professor, a Holder of the Fratamico Endowed Chair, and the Director of System Energy Efficiency Laboratory, University of California at San Diego, La Jolla, CA, USA. From 1998 to 2005, she was a full-time Research Scientist with HP Labs, Palo Alto, CA, USA, while also leading research efforts with Stanford University, Stanford, CA, USA. She was a Senior Design Engineer with Altera Corporation, San Jose, CA, USA. She is leading a number of projects, including efforts funded by DARPA/SRC JUMP 2.0 PRISM program with focus on design of accelerators for analysis of big data, DARPA and NSF funded projects on hyperdimensional computing, and SRC funded project on IoT system reliability and maintainability. Her current research interests include energy-efficient computing, cyber–physical, and distributed systems.
\end{IEEEbiography}

\end{document}